\documentclass[onecolumn]{IEEEtran}
\usepackage{lmodern}
\usepackage[T1]{fontenc}
\usepackage[latin9]{inputenc}
\usepackage{float}
\usepackage{amsmath}
\usepackage{amsthm}
\usepackage{amssymb}
\usepackage{graphicx}
\usepackage[unicode=true,
 bookmarks=true,bookmarksnumbered=true,bookmarksopen=true,bookmarksopenlevel=1,
 breaklinks=false,pdfborder={0 0 0},pdfborderstyle={},backref=false,colorlinks=false]
 {hyperref}
\hypersetup{
 hypertexnames=false}

\makeatletter

\floatstyle{ruled}
\newfloat{algorithm}{tbp}{loa}
\providecommand{\algorithmname}{Algorithm}
\floatname{algorithm}{\protect\algorithmname}

\theoremstyle{plain}
\newtheorem{thm}{\protect\theoremname}[section]
\theoremstyle{definition}
\newtheorem{defn}[thm]{\protect\definitionname}
\theoremstyle{plain}
\newtheorem{prop}[thm]{\protect\propositionname}
\theoremstyle{plain}
\newtheorem{cor}[thm]{\protect\corollaryname}
\theoremstyle{plain}
\newtheorem{lem}[thm]{\protect\lemmaname}

\@ifundefined{date}{}{\date{}}
\usepackage{subcaption}

\makeatother

\providecommand{\corollaryname}{Corollary}
\providecommand{\definitionname}{Definition}
\providecommand{\lemmaname}{Lemma}
\providecommand{\propositionname}{Proposition}
\providecommand{\theoremname}{Theorem}

\begin{document}

\title{Game of Thrones: Fully Distributed Learning for Multi-Player Bandits}

\author{Ilai Bistritz and Amir Leshem\thanks{Ilai Bistritz is with the department of Electrical Engineering, Stanford
University, USA, e-mail: bistritz@stanford.edu.\protect \\
Amir Leshem is with the Faculty of Engineering, Bar-Ilan University,
Ramat-Gan, Israel, e-mail: leshema@biu.ac.il. \protect \\
This research was supported by the Israeli Ministry of Science and
Technology under grant 3-13038 and by a joint ISF-NRF research grant
number 2277/16. A paper with a preliminary version of this study (\cite{bistritz2018distributed})
was published in the thirty-second Conference on Neural Information
Processing Systems (Neurips 2018).}}
\maketitle
\begin{abstract}
We consider an $N$-player multi-armed bandit game where each player
chooses one out of $M$ arms for $T$ turns. Each player has different
expected rewards for the arms, and the instantaneous rewards are independent
and identically distributed or Markovian. When two or more players
choose the same arm, they all receive zero reward. Performance is
measured using the expected sum of regrets, compared with an optimal
assignment of arms to players that maximizes the sum of expected rewards.
We assume that each player only knows her actions and the reward she
received each turn. Players cannot observe the actions of other players,
and no communication between players is possible. We present a distributed
algorithm and prove that it achieves an expected sum of regrets of
near-$O\left(\log T\right)$. This is the first algorithm to achieve
a near order optimal regret in this fully distributed scenario. All
other works have assumed that either all players have the same vector
of expected rewards or that communication between players is possible.
 
\end{abstract}

\section{Introduction }

In online learning problems, an agent needs to learn on the run how
to behave optimally. The crux of these problems is the tradeoff between
exploration and exploitation. This tradeoff is well captured by the
multi-armed bandit problem, which has attracted enormous attention
from the research community \cite{Audibert2013,Cesa-Bianchi2006,Denardo2007,Krishnamurthy2009,Russo2014,Yu2009,Agrawal}.
In the multi-armed bandit problem, on each turn, for a total of $T$
turns, an agent has to choose to pull one of the arms of several slot
machines (bandits). Each arm provides a stochastic reward with a distribution
that is unknown to the agent. The agent's performance is measured
by the expected difference between the sum of rewards and the sum
of rewards she could have achieved if she knew the statistics of the
machines. In their seminal work, \cite{Lai1985} proved that the best
policy achieves a regret of $O\left(\log T\right)$. 

Recently, there has been growing interest in the case of the multi-player
multi-armed bandit. In the multi-player scenario, the nature of the
interaction between the players can take many forms. Players may want
to solve the problem of finding the best arms as a team \cite{Shahrampour2017,Hillel2013,Landgren2016,Cesa-Bianchi2016,Korda2016,Szorenyi2013,dimakopoulou2018coordinated},
or may compete over the arms as resources they all individually require
\cite{Rosenski2016,Nayyar2016,Kalathil2014,Liu2010,Vakili2013,Lai2008,Anandkumar2011,Liu2013,Avner2014,evirgen2017effect}.
The idea of regret in the competitive multi-player multi-armed bandit
problem is the expected sum of regrets and is defined as the performance
loss compared to the optimal assignment of arms to players. The rationale
for this notion of regret is formulated from the designer's perspective,
who wants the distributed system of individuals to converge to a globally
good solution. 

Many works have considered a scenario where all the players have the
same expectations for the rewards of all arms. Some of these works
assume that communication between players is possible \cite{Liu2013,Vakili2013,Lai2008,Liu2010},
whereas others consider a fully distributed scenario \cite{Rosenski2016,Anandkumar2011,Avner2014}.

One of the main reasons for studying resource allocation bandits has
to do with their applications in wireless networks. In these scenarios,
the channels are interpreted as the arms and the channel gain (or
signal to noise ratio) as the arm's reward. However, since users are
scattered in space, the physical reality dictates that different arms
have different expected channel gains for different players. Different
users having different preferences is of course the typical case in
many other resource allocation scenarios as well.

This essential generalization for a matrix of expected rewards introduces
the famous assignment problem \cite{Papadimitriou98}. Achieving a
sublinear expected total regret in a distributed manner requires a
distributed solution to the assignment problem, which has been explored
in \cite{Bertsekas1988,Zavlanos2008,naparstek2013fully}. This generalization
was first considered in \cite{Kalathil2014}, and later enhanced in
\cite{Nayyar2016}, where an algorithm that achieves an expected sum
of regrets of near-$O\left(\log T\right)$ was presented. However,
this algorithm requires communication between players. It is based
on the distributed auction algorithm in \cite{Bertsekas1988}, which
is not fully distributed. It requires that players can observe the
bids of other players. This was possible in \cite{Nayyar2016,Kalathil2014}
since it was assumed that the players could observe the actions of
other players, which allowed them to communicate by using arm choices
as a signaling method. The work in \cite{Avner2016} suggested an
algorithm that only assumes users can sense all channels without knowing
which channels were chosen by whom. This algorithm requires less communication
than \cite{Nayyar2016}, but has no regret guarantees. If the multi-armed
bandit problem is relaxed such that the expected rewards are multiplies
of some constant and players can choose not to transmit but instead
only to sense a chosen channel, then it is shown in \cite{Zafaruddin2019}
that a regret of $O\left(\log T\right)$ can be achieved. 

In wireless networks, assuming that each user can hear all the other
transmissions (fully connected network) is very demanding in practice.
It requires a large sensing overhead or might simply be impossible
due to the geometry of the network (e.g., exposed and hidden terminals).
In a fully distributed scenario, players only have access to their
previous actions and rewards. However, to date, there is no completely
distributed algorithm that converges to the exact optimal solution
of the assignment problem. The fully distributed multi-armed bandit
problem remains unresolved. 

Our work generalizes \cite{Rosenski2016} for different expectations
for different players and \cite{Nayyar2016,Kalathil2014,Avner2016}
for a fully distributed scenario with no communication between players. 

Recently, powerful payoff-based dynamics were introduced by \cite{Pradelski2012,Marden2014}.
These dynamics only require each player to know her own action and
the reward she received for that action. The dynamics in \cite{Pradelski2012}
guarantee that the Nash equilibrium (NE) with the best sum of utilities
strategy profile will be played a sufficiently large portion of the
time. The dynamics in \cite{Marden2014} guarantee that the optimal
sum of utilities strategy profile will be played a sufficiently large
portion of the time, even if it is not a NE. In \cite{Menon2013},
equipping the dynamics of \cite{Marden2014} with a decreasing exploration
rate sequence was shown to provide a convergence in probability guarantee
to the optimal sum of utilities solution. However, no explicit probability
of convergence in a specific finite time was provided, which is essential
for regret computation. Nevertheless, the crucial issue of applying
these results to our case is that they all assume interdependent games.
Interdependent games are games where each group of players can always
influence at least one player from outside this group. In the multiplayer
multi-armed bandit collision model, this does not hold. A player who
shares an arm with another receives zero reward. Nothing that other
players (who chose other arms) can do will change this. 

In this paper, we suggest novel modified dynamics that behave similarly
to \cite{Marden2014}, but in our non-interdependent game. These dynamics
guarantee that the optimal solution to the assignment problem is played
a considerable amount of the time. We present a fully distributed
multi-player multi-armed bandit algorithm for the resource allocation
and collision scenario, based on these modified dynamics. By fully
distributed we mean that players only have access to their own actions
and rewards. To the best of our knowledge, this is the first algorithm
that achieves a near-optimal ($O\left(\log T\right)$) expected sum
of regrets with a matrix of expected rewards and no communication
at all between players, or equivalently, a game where players cannot
observe the actions of other players. 

A preliminary version of this paper appeared in \cite{bistritz2018distributed}.
The algorithm and analysis in this paper substantially extend the
results as follows:
\begin{enumerate}
\item The total regret is improved to near-$O\left(\log T\right)$ instead
of near-$O\left(\log^{2}T\right)$ of \cite{bistritz2018distributed},
(almost) closing the gap with the lower bound of $O\left(\log T\right)$. 
\item A tighter stationary distribution analysis of the GoT phase (in Appendix
A) allows for choosing a much better parameter $c$ for the algorithm,
improving the convergence time significantly. 
\item The stochastic arm rewards can be unbounded or Markovian, and not
only bounded i.i.d. variables as in \cite{bistritz2018distributed}. 
\end{enumerate}

\subsection{Outline}

Section 2 formalizes the multi-player multi-armed bandit resource
allocation game. Section 3 describes our fully distributed Game of
Thrones (GoT) algorithm. In the first phase of every epoch, players
explore in order to estimate the expectations of the arm rewards.
This phase is analyzed in Section 4. In the second phase of every
epoch, players use our modified dynamics and play the optimal solution
most of the time. This is analyzed in Section 5. In the third and
final phase of every epoch, players play the action they played most
of the time in the recent GoT phases. Section 6 generalizes our main
result to the case of Markovian rewards. Section 7 demonstrates our
algorithm's performance using numerical experiments and Section 8
concludes the paper. 

\section{Problem Formulation }

We consider a stochastic game with the set of players $\mathcal{N}=\left\{ 1,...,N\right\} $
and a finite time horizon $T$. The horizon $T$ is not known in advance
by any of the players. The discrete turn index is denoted by $t$.
The strategy space of each player is a set of $M$ arms denoted by
$\mathcal{A}_{n}=\left\{ 1,...,M\right\} $ for each $n$. We assume
that $M\geq N$, such that an allocation without collisions is possible.
At each turn $t$, all players simultaneously pick one arm each. The
arm that player $n$ chooses at time $t$ is denoted by $a_{n}\left(t\right)$
and the strategy profile at time $t$ is $\boldsymbol{a}\left(t\right)$.
Players do not know which arms the other players chose, and need not
even know how many other players there are.

Define the set of players that chose arm $i$ in strategy profile
$\boldsymbol{a}$ as $\mathcal{N}_{i}\left(\boldsymbol{a}\right)=\left\{ n\,|\,a_{n}=i\right\} .$
The no-collision indicator of arm $i$ in strategy profile $\boldsymbol{a}$
is defined as
\begin{equation}
\eta_{i}\left(\boldsymbol{a}\right)=\Biggl\{\begin{array}{cc}
0 & \Bigl|\mathcal{N}_{i}\left(\boldsymbol{a}\right)\Bigr|>1\\
1 & o.w.
\end{array}.\label{eq:2}
\end{equation}
The assumptions on the stochastic rewards are summarized as follows.
We also study the case of Markovian rewards in Section \ref{sec:Markovian-Rewards}.
\begin{defn}
\label{Def: i.i.d rewards}The sequence of rewards $\left\{ r_{n,i}\left(t\right)\right\} _{t=1}^{T}$
of arm $i$ for player $n$ is i.i.d. with expectation $\mu_{n,i}>0$
and variance $\sigma_{n,i}^{2}$ such that:
\begin{enumerate}
\item The distribution of $r_{n,i}\left(t\right)$ is continuous for each
$n,i$.
\item For some positive parameter $b_{n,i}$ we have $\mathbb{E}\left\{ \left|r_{n,i}\left(t\right)-\mu_{n,i}\right|^{k}\right\} \leq\frac{1}{2}k!\sigma_{n,i}^{2}b_{n,i}^{k-2}$
for all integers $k\geq3$.
\item The sequences $\left\{ r_{n,i}\left(t\right)\right\} _{t}$ are independent
for different $n$ or different $i$.
\end{enumerate}
\end{defn}
The family of distributions that satisfies the second condition (known
as Bernstein's condition, see \cite{1937}) includes, among many others,
the normal and Laplace distributions and, trivially, any bounded distribution.
The instantaneous utility of player $n$ in strategy profile $\boldsymbol{a}\left(t\right)$
at time $t$ is
\begin{equation}
\upsilon_{n}\left(\boldsymbol{a}\left(t\right)\right)=r_{n,a_{n}\left(t\right)}\left(t\right)\eta_{a_{n}\left(t\right)}\left(\boldsymbol{a}\left(t\right)\right).\label{eq:3}
\end{equation}
Our goal is to design a distributed algorithm that minimizes the expected
total regret, defined next. 
\begin{defn}
Denote the expected utility of player $n$ in the strategy profile
$\boldsymbol{a}$ by $g_{n}\left(\boldsymbol{a}\right)=\mathbb{E}\left\{ \upsilon_{n}\left(\boldsymbol{a}\right)\right\} $.
The total regret is defined as the random variable 
\begin{equation}
R=\sum_{t=1}^{T}\sum_{n=1}^{N}\upsilon_{n}\left(\boldsymbol{a}^{*}\right)-\sum_{t=1}^{T}\sum_{n=1}^{N}r_{n,a_{n}\left(t\right)}\left(t\right)\eta_{a_{n}\left(t\right)}\left(\boldsymbol{a}\left(t\right)\right)\label{eq:4}
\end{equation}
where
\begin{equation}
\boldsymbol{a}^{*}\in\arg\underset{\boldsymbol{a}}{\max}\sum_{n=1}^{N}g_{n}\left(\boldsymbol{a}\right).\label{eq:5}
\end{equation}
The expected total regret $\bar{R}\triangleq\mathbb{E}\left\{ R\right\} $
is the average of \eqref{eq:4} over the randomness of the rewards
$\left\{ r_{n,i}\left(t\right)\right\} _{t}$, that dictate the random
actions $\left\{ a_{n}\left(t\right)\right\} $. 
\end{defn}
The problem in \eqref{eq:5} is no other than the famous assignment
problem \cite{Papadimitriou98} on the $N\times M$ matrix of expectations
$\left\{ \mu_{n,i}\right\} $. In this sense, our problem is a generalization
of the distributed assignment problem to an online learning framework.

Assuming continuously distributed rewards is well justified in wireless
networks. Given no collision, the quality of an arm (channel) always
has a continuous measure like SNR or a channel gain. Since the probability
of zero reward in a non-collision is zero, players can safely deduce
their non-collision indicator and rule out collisions in their estimation
of the expected rewards. In Section \ref{sec:Markovian-Rewards},
we extend our results to discrete Markovian rewards, that include
i.i.d. discrete rewards as a special case. This reflects a case where
each user can operate in one out of a finite number of qualities of
service.

In the case where the probability for a zero reward is not zero, we
can assume instead that each player can observe her collision indicator
in addition to her reward. Knowing whether other players chose the
same arm is a very modest requirement compared to assuming that players
can observe the actions of other players. 

According to the seminal work \cite{Lai1985}, the optimal regret
of the single-player case is logarithmic; i.e., $O\left(\log T\right)$.
The multiple players do not help each other; hence, we expect the
expected total regret lower bound to be logarithmic at best, as shown
by the following proposition. 
\begin{prop}
The expected total regret is at least $\Omega\left(\log T\right)$.
\end{prop}
\begin{IEEEproof}
Assume that for $N>1$ there is a policy that results in a total expected
regret lower than $\Omega\left(\log T\right)$. Some player with given
expected rewards, denoted player $n$, can simulate $N-1$ other players
and set their expected rewards such that she gets her best arm at
the optimal matching, and all other players have the same expected
reward for all arms. In this case, her personal term in the sum of
regrets in \eqref{eq:4} is at least her single player regret (can
be larger due to collisions), and is lower than the total sum which
is lower than $\Omega\left(\log T\right)$. This player can generate
the rewards of other players at random, all of which are independent
of the actual rewards she receives. This player also simulates the
policies for other players, and even knows when a collision occurred
for herself and can assign zero reward in that case. Hence, simulating
$N-1$ fictitious players is a valid single player multi-armed bandit
policy that violates the $\Omega\left(\log T\right)$ bound, which
is a contradiction. We conclude that this bound is also valid for
$N>1$.
\end{IEEEproof}

\section{Game of Thrones Algorithm}

When all players have the same arm expectations, the exploration phase
is used to identify the $N$ best arms. Once the best arms are identified,
players need to coordinate to be sure that each of them will sit on
a different ``chair'' (see the Musical Chairs algorithm in \cite{Rosenski2016}).
When players have different arm expectations, a non-cooperative \textbf{game}
is induced where the estimated expected rewards serve as utilities.
In this game, players cannot sit on an ordinary chair without causing
a linear regret, and must strive for a single \textbf{throne}. This
throne is the arm they must play in the allocation that maximizes
the sum of the expected rewards of all players. Any other solution
will result in linear (in $T$) expected total regret. We assume that
our assignment problem has a unique optimal allocation which occurs
with probability 1 if the expected rewards are generated at random
using a continuous distribution (i.e., \textquotedblleft for almost
all games\textquotedblright ).

\begin{figure*}[t]
~~~~~~~~~~~~~~~~~~~~~\includegraphics[width=13cm,height=1cm]{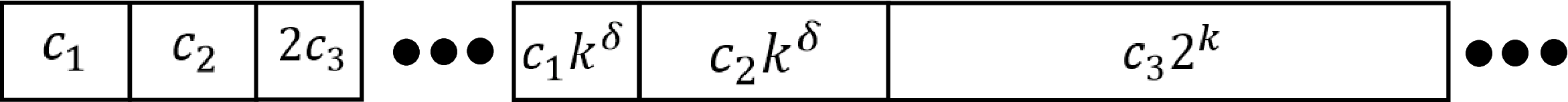}

\caption{Epoch structure. Depicted are the first and the $k$-th epochs. \label{fig:Epoches-Structure}}
\end{figure*}

The total time needed for exploration increases with $T$ since the
cost of being wrong becomes higher. When $T$ is known by the players,
a long enough exploration can be accomplished at the beginning of
the game. In order to maintain the right balance between exploration
and exploitation when $T$ is not known in advance to the players,
we divide the $T$ turns into epochs, one starting immediately after
the other. Each epoch is further divided into three phases. In the
$k$-th epoch:
\begin{enumerate}
\item \textbf{Exploration Phase} - this phase has a length of $\left\lceil c_{1}k^{\delta}\right\rceil $
turns for an arbitrary positive integer $c_{1}$ and a positive $\delta$.
The goal of this phase is to estimate the expectation of the arms.
This phase is described in detail and analyzed in Section \ref{sec:Pure-Exploration-Phase}.
It adds a $O\left(\log^{1+\delta}T\right)$ to the expected total
regret. 
\item \textbf{Game of Thrones (GoT) Phase} - this phase has a length of
$\left\lceil c_{2}k^{\delta}\right\rceil $ turns for an arbitrary
positive integer $c_{2}$ and a positive $\delta$. In this phase,
players play a non-cooperative game with the estimated expectations
from the exploration phase as their deterministic utilities. They
choose their action at random according to payoff-based dynamics that
always assign a positive probability for exploration. These dynamics
induce a perturbed (ergodic) Markov chain that tends to visit the
optimal sum of utilities (in \eqref{eq:5}) state more often than
other states. When the exploration rate is low enough, the optimal
state has a probability greater than $\frac{1}{2}$ in the stationary
distribution of the chain. Hence, all players play the optimal action
most of the time and can agree on the optimal state distributedly.
These arguments are described in detail and analyzed in Section \ref{sec:Game-of-Thrones}.
This phase adds a $O\left(\log^{1+\delta}T\right)$ to the expected
total regret. The GoT dynamics are described in the next subsection.
\item \textbf{Exploitation Phase} - this phase has a length of $c_{3}2^{k}$
turns for an arbitrary positive integer $c_{3}$. During this phase,
each player plays the most frequent action she has played during the
last $\left\lfloor \frac{k}{2}\right\rfloor +1$ GoT phases combined.
Since the error probability of the exploration phase is kept small,
it is very likely that all the last $\left\lfloor \frac{k}{2}\right\rfloor +1$
exploration phases agree on the same optimal state. Hence, by playing
their most frequent action, players are highly likely to play the
optimal strategy profile that has a stationary distribution of more
than $\frac{1}{2}$ in all the last $\left\lfloor \frac{k}{2}\right\rfloor +1$
GoT Markov chains. This phase adds a vanishing term (with $T$) to
the expected total regret.
\end{enumerate}
This division into epochs is depicted in Fig. \ref{fig:Epoches-Structure}.
The GoT algorithm is described in Algorithm \ref{alg:Game-of-Thrones}.
We note that the constants $c_{1},c_{2},c_{3}$ can be arbitrarily
chosen and do not need to satisfy any condition. Their only role is
to adjust the length of the initial three phases, which is only of
interest for numerical or practical reasons (see Section \ref{sec:Simulation-Results}).

\subsection{Game of Thrones Dynamics\label{subsec:Game-of-Thrones-Dynamics}}

The core of our algorithm is the GoT dynamics of the second phase,
where players play the game of thrones with the utility function $u_{n}\left(\boldsymbol{a}\right)$
for player $n$. Denote the optimal objective by $J_{1}$ and $J_{0}=\sum_{n}u_{n,\max}$
for $u_{n,\max}=\underset{\boldsymbol{a}}{\max}\,u_{n}\left(\boldsymbol{a}\right)$.
Let  $c\geq J_{0}-J_{1}$. Each player $n$ has a personal state $S_{n}=\left\{ C,D\right\} $
where $C$ is content and $D$ is discontent. Each player keeps a
baseline action $\overline{a}_{n}$. In each turn during the GoT phase:
\begin{itemize}
\item A content player has a very small probability of deviating from her
current baseline action:
\begin{equation}
p_{n}^{a_{n}}=\Biggl\{\begin{array}{cc}
\frac{\varepsilon^{c}}{\Bigl|\mathcal{A}_{n}\Bigr|-1} & a_{n}\neq\overline{a}_{n}\\
1-\varepsilon^{c} & a_{n}=\overline{a}_{n}
\end{array}.\label{eq:6}
\end{equation}
\item A discontent player chooses an action uniformly at random; i.e., 
\begin{equation}
p_{n}^{a_{n}}=\frac{1}{\Bigl|\mathcal{A}_{n}\Bigr|},\,\forall a_{n}\in\mathcal{A}_{n}.\label{eq:7}
\end{equation}
\end{itemize}
The transitions between $C$ and $D$ are determined as follows:
\begin{itemize}
\item If $\overline{a}_{n}=a_{n}$ and $u_{n}>0$, then a content player
remains content with probability 1:
\begin{equation}
\left[\overline{a}_{n},C\right]\rightarrow\left[\overline{a}_{n},C\right].\label{eq:8}
\end{equation}
\item If $\overline{a}_{n}\neq a_{n}$ or $u_{n}=0$ or $S_{n}=D$, then
the state transition is ($C/D$ denoting either $C$ or $D$):
\begin{equation}
\left[\overline{a}_{n},C/D\right]\rightarrow\Biggl\{\begin{array}{cc}
\left[a_{n},C\right] & \frac{u_{n}}{u_{n,\max}}\varepsilon^{u_{n,\max}-u_{n}}\\
\left[a_{n},D\right] & 1-\frac{u_{n}}{u_{n,\max}}\varepsilon^{u_{n,\max}-u_{n}}
\end{array}.\label{eq:9}
\end{equation}
\end{itemize}
We now formulate our main result on the regret of our algorithm, which
is generalized to Markovian rewards in Section \ref{sec:Markovian-Rewards}.
The proof of Theorem \ref{thm:StationaryBound} provides an explicit
requirement on the parameter $\varepsilon$. 
\begin{thm}[Main Theorem]
\label{thm:Main}Assume i.i.d. rewards $\left\{ r_{n,i}\left(t\right)\right\} _{t}$
with positive expectations $\left\{ \mu_{n,i}\right\} $ as in Definition
\ref{Def: i.i.d rewards}, such that \eqref{eq:5} has a unique solution.
Let the game have a finite horizon $T$, unknown to the players. Let
each player play according to Algorithm \ref{alg:Game-of-Thrones},
with any integers $c_{1},c_{2},c_{3}>0$ and $\delta>0$. Then there
exists a small enough $\varepsilon$ such that for large enough $T$,
the expected total regret is bounded by
\begin{equation}
\bar{R}\leq4\left(\underset{n,i}{\max}\mu_{n,i}\right)\left(c_{1}+c_{2}\right)N\log_{2}^{1+\delta}\left(\frac{T}{c_{3}}+2\right)=O\left(\log^{1+\delta}T\right).\label{eq:12}
\end{equation}
\end{thm}
\begin{IEEEproof}
Let $\delta>0$. Let $k_{0}$ be the index of a sufficiently large
epoch. We now bound the expected total regret of epoch $k>k_{0}$,
denoted by $\bar{R}_{k}$. Define $E_{k}$ as the event where the
$k$-th exploitation phase does not consist of playing the optimal
assignment $\boldsymbol{a}^{*}$. We have 
\begin{equation}
\mathbb{P}\left(E_{k}\right)\leq\mathbb{P}\left(\bigcup_{r=0}^{\left\lfloor \frac{k}{2}\right\rfloor }P_{e,k-r}\right)+P_{c,k}\label{eq:15}
\end{equation}
where $P_{e,k}$ bounds the probability that the $k$-th exploration
phase results in an estimation for which the optimal assignment is
different from $\boldsymbol{a}^{*}$ (see Lemma \ref{lem:Percision}
and \eqref{eq:23}), and $P_{c,k}$ is the probability that $\boldsymbol{a}^{*}$
was not the most frequent state in the last $\left\lfloor \frac{k}{2}\right\rfloor +1$
GoT phases combined, given that the $\left\lfloor \frac{k}{2}\right\rfloor $
recent exploration phases succeeded.  If neither of these failures
occurred, then $\boldsymbol{a}^{*}$ is played in the $k$-th exploitation
phase, which establishes \eqref{eq:15}. Only under $E_{k}$ will
this exploitation phase contribute to the total regret. Let $\mu_{\max}=\underset{n,i}{\max}\mu_{n,i}$
. For $k>k_{0}$ we obtain, for any $0<\eta<\frac{1}{2}$, that
\begin{multline}
\bar{R}_{k}\leq\mu_{\max}\left(\left(c_{1}+c_{2}\right)k^{\delta}+2\right)N+\left(\mathbb{P}\left(\bigcup_{r=0}^{\left\lfloor \frac{k}{2}\right\rfloor }P_{e,k-r}\right)+P_{c,k}\right)\mu_{\max}c_{3}2^{k}N\underset{\left(a\right)}{\leq}\mu_{\max}\left(\left(c_{1}+c_{2}\right)k^{\delta}+2\right)N+\\
\left(\frac{2NM}{1-e^{-wc_{1}\left(\frac{k}{4}\right)^{\delta}}}e^{-\frac{w}{2}c_{1}\left(\frac{k}{4}\right)^{\delta}k}+\frac{NM}{1-e^{-\frac{1}{36M^{2}}c_{1}\left(\frac{k}{4}\right)^{\delta}}}e^{-\frac{1}{72M^{2}}c_{1}\left(\frac{k}{4}\right)^{\delta}k}+\left(C_{0}e^{-\frac{c_{2}\eta^{2}}{144T_{m}\left(\frac{1}{8}\right)}\left(\pi_{z^{*}}-\frac{1}{2\left(1-\eta\right)}\right)\left(\frac{k}{2}\right)^{\delta}}\right)^{k}\right)\mu_{\max}c_{3}2^{k}N\\
\underset{\left(b\right)}{\leq}2\mu_{\max}\left(c_{1}+c_{2}\right)k^{\delta}N\label{eq:16}
\end{multline}
where (a) is due to the upper bounds on $\mathbb{P}\left(\bigcup_{r=0}^{\left\lfloor \frac{k}{2}\right\rfloor }P_{e,k-r}\right),\,P_{c,k}$
from Lemma \ref{lem: exploration} and Lemma \ref{lem:ConnectingGoTs},
respectively, where $w$ is a positive constant defined in \eqref{eq:22}.
Given that the $\left\lfloor \frac{k}{2}\right\rfloor $ recent exploration
phases succeeded, using Theorem \ref{thm:StationaryBound} (see \eqref{eq:68},\eqref{eq:69})
we deduce that there exists a small enough $\varepsilon$ such that
Lemma \ref{lem:ConnectingGoTs} holds with $\pi_{z^{*}}=\underset{k-\left\lfloor \frac{k}{2}\right\rfloor \leq i\leq k}{\min}\pi_{z^{i*}}>\frac{1}{2\left(1-\eta\right)}$,
where $\pi_{z^{i*},}$ is the component of the optimal state $z^{*}$
(of $\boldsymbol{a}^{*}$) in the stationary distribution of the Markov
chain of the $i$-th GoT phase. Inequality (b) follows since for $k>k_{0}$
we have 
\begin{equation}
\max\left\{ e^{-\frac{w}{2}c_{1}\left(\frac{k}{4}\right)^{\delta}},\,e^{-\frac{1}{72M^{2}}c_{1}\left(\frac{k}{4}\right)^{\delta}},\,C_{0}e^{-\frac{c_{2}\eta^{2}}{144T_{m}\left(\frac{1}{8}\right)}\left(\pi_{z^{*}}-\frac{1}{2\left(1-\eta\right)}\right)\left(\frac{k}{2}\right)^{\delta}}\right\} <\frac{1}{2}\label{eq:17}
\end{equation}
for $T_{m}\left(\frac{1}{8}\right)=\underset{k-\left\lfloor \frac{k}{2}\right\rfloor \leq i\leq k}{\max}T_{m,i}\left(\frac{1}{8}\right)$,
where $T_{m,i}\left(\frac{1}{8}\right)$ is the mixing time of the
Markov chain of the $i$-th GoT phase\footnote{From any state $z$, there is a probability of at least $\left(\frac{\varepsilon^{c}}{K-1}\right)^{N}$
that the next state is $z_{0}=\left[\boldsymbol{1}_{N},D^{N}\right]$.
Then after $T_{u}$ turns the probability that two independent instances
of $Z$ do not meet at $\boldsymbol{z}_{0}$ is less than $\left(1-\left(\frac{\varepsilon^{c}}{K-1}\right)^{2N}\right)^{T_{u}}$,
so the coupling Lemma yields $T_{m}\left(\frac{1}{8}\right)\leq\frac{\ln8}{\left(\frac{\varepsilon^{c}}{K-1}\right)^{2N}}<\infty$.}. Let $E$ be the number of epochs that start within $T$ turns. We
conclude that
\begin{equation}
\bar{R}\underset{(a)}{\leq}\sum_{k=1}^{E}\bar{R}_{k}\underset{(b)}{\leq}N\mu_{\max}\sum_{k=1}^{k_{0}}c_{3}2^{k}+2N\mu_{\max}\sum_{k=1}^{E}\left(c_{1}+c_{2}\right)k^{\delta}\underset{(c)}{\leq}N\mu_{\max}c_{3}2^{k_{0}+1}+2\mu_{\max}\left(c_{1}+c_{2}\right)N\log_{2}^{1+\delta}\left(\frac{T}{c_{3}}+2\right)\label{eq:18}
\end{equation}
where (a) follows since completing the last epoch to be a full epoch
only increases $\bar{R}_{k}$. In (b) we used \eqref{eq:16} for $k>k_{0}$
and $\bar{R}_{k}\leq2\mu_{\max}N\left(c_{1}+c_{2}\right)k^{\delta}+\mu_{\max}Nc_{3}2^{k}$
for $k\leq k_{0}$. Inequality (c) follows from $\sum_{k=k_{0}+1}^{E}k^{\delta}\leq E^{1+\delta}$
and bounding $E$ using $T\geq\sum_{k=1}^{E-1}c_{3}2^{k}\geq c_{3}\left(2^{E}-2\right)$. 
\end{IEEEproof}
\begin{algorithm}[tbh]
\caption{\label{alg:Game-of-Thrones}Game of Thrones Algorithm}
\textbf{Initialization} -Set $\delta>0$, $\varepsilon>0$ and integers
$c_{1},c_{2},c_{3}>0$. Set $V_{n,i}\left(0\right)=0$, $s_{n,i}\left(0\right)=0$
for each arm $i=1,..,M$.

\textbf{For $k=1,...,E$ }
\begin{enumerate}
\item \textbf{Exploration Phase} - for the next $\left\lceil c_{1}k^{\delta}\right\rceil $
turns
\begin{enumerate}
\item Sample an arm $i$ uniformly at random from all $M$ arms. 
\item Receive the reward $r_{n,i}\left(t\right)$ and set $\eta_{i}\left(t\right)=0$
if $r_{n,i}\left(t\right)=0$ and $\eta_{i}\left(t\right)=1$ otherwise. 
\item If $\eta_{i}\left(t\right)=1$ then update $V_{n,i}\left(t\right)=V_{n,i}\left(t-1\right)+1$
and $s_{n,i}\left(t\right)=s_{n,i}\left(t-1\right)+r_{n,i}\left(t\right)$.
\item Estimate the expectation of arm $\text{\ensuremath{i} by \ensuremath{\mu_{n,i}^{k}=\frac{s_{n,i}\left(t\right)}{V_{n,i}\left(t\right)}}}$,
for each $i=1,...,M$.
\end{enumerate}
\item \textbf{GoT Phase -} for the next $\left\lceil c_{2}k^{\delta}\right\rceil $
turns, play according to the GoT Dynamics with $\varepsilon$. Set
$S_{n}=C$ and set $\overline{a}_{n}$ to the last action played in
the $k-\left\lfloor \frac{k}{2}\right\rfloor -1$ GoT phase, or a
random action if $k=1,2$.
\begin{enumerate}
\item If $S_{n}=C$ then play according to \eqref{eq:6} and if $S_{n}=D$
then choose an arm at random \eqref{eq:7}.
\item If $\overline{a}_{n}\neq a_{n}$ or $u_{n}=0$ or $S_{n}=D$ then
set $S_{n}=C$ w.p. $\frac{u_{n}}{u_{n,\max}}\varepsilon^{u_{n,\max}-u_{n}}$,
and otherwise set $S_{n}=D$.
\item Keep track of the number of times each action was played and resulted
in being content:
\begin{equation}
F_{k}^{n}\left(i\right)\triangleq\sum_{t\in\mathcal{G}_{k}}I\left(a_{n}\left(t\right)=i,S_{n}\left(t\right)=C\right)\label{eq:10}
\end{equation}
where $I$ is the indicator function and $\mathcal{G}_{k}$ is the
set of turns of the $k$-th GoT phase. 
\end{enumerate}
\item \textbf{Exploitation Phase} - for the next $c_{3}2^{k}$ turns, play
\begin{equation}
a_{n}^{k}=\arg\underset{i=1,...,M}{\max}\sum_{r=0}^{\left\lfloor \frac{k}{2}\right\rfloor }F_{k-r}^{n}\left(i\right)\label{eq:11}
\end{equation}
\end{enumerate}
\textbf{End}
\end{algorithm}

The goal of the parameter $\delta>0$ is to guarantee that Algorithm
\ref{alg:Game-of-Thrones} can work without the knowledge of the problem
parameters $w,T_{m}\left(\frac{1}{8}\right)$ and $\pi_{z^{*}}$ (see
\eqref{eq:17}). Since the $k$-th exploration and GoT phases both
have an increasing length of $k^{\delta}$, they are guaranteed to
eventually be long enough compared to any of these parameters. The
cost is achieving only a near-optimal regret of $O\left(\log^{1+\delta}T\right)$
instead of $O\left(\log T\right)$. If bounds on these parameters
are available, it is possible to choose $c_{1}$ and $c_{2}$ such
that \eqref{eq:17} holds even for $\delta=0$.

If either the exploration or the GoT phases fail, the regret becomes
linear with $T$. Like many other bandit algorithms, we avoid a linear
expected regret by ensuring that the error probabilities vanish exponentially
with $T$. By using a single epoch with a constant duration for the
first two phases, we get an alternative formulation of our result,
as in \cite{Rosenski2016}. In this case, with high probability (in
$T$) our algorithm achieves a constant regret. 
\begin{cor}
For any $\eta>0$, there exist $c_{1},c_{2}$ such that a GoT algorithm
with a single epoch (of length $T$) has a constant regret in $T$
with probability of at least $1-\eta$.
\end{cor}
\begin{IEEEproof}
This corollary follows since $P_{e,k}$ in \eqref{eq:23} vanishes
with $c_{1}$ and $P_{c,1}$ in \eqref{eq:34} vanishes with $c_{2}$. 
\end{IEEEproof}

\section{The Exploration Phase - Estimation of the Expected Rewards\label{sec:Pure-Exploration-Phase}}

In this section, we describe the exploration phase, and analyze its
addition to the expected total regret. At the beginning of the game,
players still do not have any evaluation of the $M$ different arms.
They estimate these values on the run, based on the rewards they get.
We propose a pure exploration phase where each player picks an arm
uniformly at random, similar to the one suggested in \cite{Rosenski2016}.
Note that in contrast to \cite{Rosenski2016}, we do not assume that
$T$ is known to the players. Hence, the exploration phase is repeated
in each epoch. However, the estimation uses all previous exploration
phases, so that the number of samples for estimation grows like $\varTheta\left(k^{1+\delta}\right)$
over time.

We use the following notion of the second best objective of the assignment
problem. Note that $J_{2}<J_{1}$ is in general different from the
objective of second best allocation, which might be also optimal.
If all allocations have the same objective then there is no need for
any estimation and all the results of this section trivially hold
with $c_{1}\geq1$.
\begin{defn}
Let $J_{1}=\underset{\boldsymbol{a}}{\max}\sum_{n=1}^{N}g_{n}\left(\boldsymbol{a}\right)$.
We define the second best objective $J_{2}$ as the maximal value
of $\sum_{n=1}^{N}g_{n}\left(\boldsymbol{a}\right)$ over all $\boldsymbol{a}$
such that $\sum_{n=1}^{N}g_{n}\left(\boldsymbol{a}\right)<J_{1}$. 
\end{defn}
The estimation of the expected rewards is never perfect. Hence, the
optimal solution to the assignment problem given the estimated expectations
might be different from the optimal solution with the correct expectations.
However, if the uncertainty of the true value of each expectation
is small enough, we expect both these optimal assignments to coincide,
as formulated in the following lemma. 
\begin{lem}
\label{lem:Percision}Assume that $\left\{ \mu_{n,i}\right\} $ are
known up to an uncertainty of $\Delta$, i.e., $\left|\hat{\mu}_{n,i}-\mu_{n,i}\right|<\Delta$
for each $n$ and $i$ for some $\left\{ \hat{\mu}_{n,i}\right\} $.
Let $\boldsymbol{a}_{1}\in\arg\underset{\boldsymbol{a}}{\max}\sum_{n=1}^{N}g_{n}\left(\boldsymbol{a}\right)$
be an optimal assignment and its objective be $J_{1}=\sum_{n=1}^{N}g_{n}\left(\boldsymbol{a}_{1}\right)$.
Let the second best objective and the corresponding assignment be
$J_{2}$ and $\boldsymbol{a}_{2}$, respectively. If $\Delta\leq\frac{J_{1}-J_{2}}{2N}$
then 
\begin{equation}
\arg\underset{\boldsymbol{a}}{\max}\sum_{n=1}^{N}g_{n}\left(\boldsymbol{a}\right)\supseteq\arg\underset{\boldsymbol{a}}{\max}\sum_{n=1}^{N}\hat{\mu}_{n,a_{n}}\eta_{a_{n}}\left(\boldsymbol{a}\right)\label{eq:19}
\end{equation}
so that the optimal assignment does not change due to the uncertainty. 
\end{lem}
\begin{IEEEproof}
First note that an optimal solution must not have any collisions,
otherwise it can be improved since $M\geq N$ and the expected rewards
are positive. Hence $J_{1}=\sum_{n=1}^{N}\mu_{n,a_{1,n}}$. For all
$n$ and $i$ we have $\hat{\mu}_{n,i}=\mu_{n,i}+z_{n,i}$ such that
$\left|z_{n,i}\right|<\Delta$. In the perturbed assignment problem,
$\boldsymbol{a}_{1}$ performs at least as well as
\begin{equation}
\sum_{n=1}^{N}\hat{\mu}_{n,a_{1,n}}=\sum_{n=1}^{N}\left(\mu_{n,a_{1,n}}+z_{n,a_{1,n}}\right)>\sum_{n=1}^{N}\mu_{n,a_{1,n}}-\Delta N\label{eq:20}
\end{equation}
and any assignment $\boldsymbol{a}\neq\boldsymbol{a}_{1}$ such that
$J\left(\boldsymbol{a}\right)<J\left(\boldsymbol{a}_{1}\right)$ performs
at most as well as
\begin{equation}
\sum_{n=1}^{N}\hat{\mu}_{n,a_{n}}\eta_{a_{n}}\left(\boldsymbol{a}\right)=\sum_{n=1}^{N}\left(\mu_{n,a_{n}}+z_{n,a_{n}}\right)\eta_{a_{n}}\left(\boldsymbol{a}\right)<\sum_{n=1}^{N}\mu_{n,a_{2,n}}\eta_{a_{2,n}}\left(\boldsymbol{a}_{2}\right)+\Delta N.\label{eq:21}
\end{equation}
Therefore, if $\Delta\leq\frac{J_{1}-J_{2}}{2N}$ then $\sum_{n=1}^{N}\hat{\mu}_{n,a_{1,n}}>\sum_{n=1}^{N}\hat{\mu}_{n,a_{n}}\eta_{a_{n}}\left(\boldsymbol{a}\right)$
for every $\boldsymbol{a}\neq\boldsymbol{a}_{1}$, which gives \eqref{eq:19}.
\end{IEEEproof}
There need not be a perturbation $\left\{ z_{n,i}\right\} $ such
that $\left|z_{n,i}\right|=\frac{J_{1}-J_{2}}{2N}$ for which the
optimal allocation of the perturbed problem is different from $\boldsymbol{a}_{1}$.
It is possible only if the allocation $a_{2}$ that yields $J_{2}$
satisfies $a_{2,n}\neq a_{1,n}$ for all $n$. Therefore, by taking
into account the linear programming constraints of the assignment
problem, it is possible to achieve a looser requirement than $\Delta\leq\frac{J_{1}-J_{2}}{2N}$.
Moreover, in practice much larger random perturbations are not likely
to change the optimal assignment. 

The following lemma concludes this section by providing an upper bound
for the probability that the $k$-th exploration phase failed, as
well as for the probability that at least one of the last $\left\lfloor \frac{k}{2}\right\rfloor +1$
exploration phases failed. 
\begin{lem}[Exploration Error Probability]
\label{lem: exploration} Assume i.i.d. rewards $\left\{ r_{n,i}\left(t\right)\right\} _{t}$
as in Definition \ref{Def: i.i.d rewards} where $b_{\max}=\underset{n,i}{\max}\,b_{n,i}$
and $\sigma_{\max}=\underset{n,i}{\max}\,\sigma_{n,i}$. Let $\left\{ \mu_{n,i}^{k}\right\} $
be the estimated reward expectations using all the exploration phases
up to epoch $k$. Let $\mathcal{A}^{*}=\arg\underset{\boldsymbol{a}}{\max}\sum_{n=1}^{N}g_{n}\left(\boldsymbol{a}\right)$
and $\boldsymbol{a}^{k*}=\arg\underset{\boldsymbol{a}}{\max}\sum_{n=1}^{N}\mu_{n,a_{n}}^{k}\eta_{a_{n}}\left(\boldsymbol{a}\right)$.
Also define $J_{1}=\sum_{n=1}^{N}g_{n}\left(\boldsymbol{a}^{*}\right)$
for $\boldsymbol{a}^{*}\in\mathcal{A}^{*}$ and the second best objective
$J_{2}$. Let $\Delta_{\min}=\underset{n}{\min}\underset{\boldsymbol{a}_{1},\boldsymbol{a}_{2}\,|\,g_{n}\left(\boldsymbol{a}_{1}\right)\neq g_{n}\left(\boldsymbol{a}_{2}\right)}{\min}\left|g_{n}\left(\boldsymbol{a}_{1}\right)-g_{n}\left(\boldsymbol{a}_{2}\right)\right|$.
Define 
\begin{equation}
w\triangleq\frac{\Delta_{\min}^{2}}{80M\left(4\sigma_{\max}^{2}+b_{\max}\Delta_{\min}\right)}.\label{eq:22}
\end{equation}
If the length of the $k$-th exploration phase is $\left\lceil c_{1}k^{\delta}\right\rceil $
then after the $k$-th epoch we have
\begin{equation}
P_{e,k}\triangleq\mathbb{P}\left(\underset{n,i}{\max}\left|\mu_{n,i}^{k}-\mu_{n,i}\right|>\frac{\Delta_{\min}}{4}\right)\leq2NMe^{-wc_{1}\left(\frac{k}{2}\right)^{\delta}k}+NMe^{-\frac{c_{1}\left(\frac{k}{2}\right)^{\delta}}{36M^{2}}k}.\label{eq:23}
\end{equation}
Furthermore,
\begin{equation}
\mathbb{P}\left(\bigcup_{r=0}^{\left\lfloor \frac{k}{2}\right\rfloor }P_{e,k-r}\right)\leq\frac{2NM}{1-e^{-wc_{1}\left(\frac{k}{4}\right)^{\delta}}}e^{-\frac{w}{2}c_{1}\left(\frac{k}{4}\right)^{\delta}k}+\frac{NM}{1-e^{-\frac{1}{36M^{2}}c_{1}\left(\frac{k}{4}\right)^{\delta}}}e^{-\frac{1}{72M^{2}}c_{1}\left(\frac{k}{4}\right)^{\delta}k}.\label{eq:24}
\end{equation}
\end{lem}
\begin{IEEEproof}
After the $k$-th exploration phase, the number of samples that are
used for estimating the expected rewards is
\begin{equation}
T_{e}\left(k\right)\triangleq c_{1}\sum_{i=1}^{k}i^{\delta}\geq c_{1}\left(\frac{k}{2}\right)^{\delta+1}.\label{eq:25}
\end{equation}
Define $A_{n,i}\left(t\right)$ as the indicator that is equal to
one if only player $n$ chose arm $i$ at time $t$. Define $\mathcal{G}_{n,i}^{A}$
as set of times for which $A_{n,i}\left(t\right)=1$. Define $V_{n,i}\left(t\right)\triangleq\sum_{\tau\in\mathcal{G}_{n,i}^{A}}A_{n,i}\left(\tau\right)$,
which is the number of visits of player $n$ to arm $i$ with no collision,
up to time $t$ and $V_{\min}=\underset{n,i}{\min}V_{n,i}\left(t\right)$.
From Definition \ref{Def: i.i.d rewards}, for each $n,i$ and for
some positive parameter $b_{n,i}$ we have $\mathbb{E}\left\{ \left|r_{n,i}\left(t\right)-\mu_{n,i}\right|^{k}\right\} \leq\frac{1}{2}k!\sigma_{n,i}^{2}b_{n,i}^{k-2}$
for all integers $k\geq3$. Define $E$ as the event in which there
exists a player $n$ that has an estimate of some arm $i$ with an
accuracy worse than $\Delta$. We have
\begin{multline}
\mathbb{P}\left(E|V_{\min}\geq v\right)=\mathbb{P}\left(\bigcup_{i=1}^{M}\bigcup_{n=1}^{N}\left\{ \left|\frac{1}{V_{n,i}\left(t\right)}\sum_{\tau\in\mathcal{G}_{n,i}^{A}}r_{n,i}\left(\tau\right)-\mu_{n,i}\right|>\Delta\,|\,V_{\min}\geq v\right\} \right)\\
\underset{\left(a\right)}{\leq}NM\underset{n,i}{\max}\mathbb{P}\left(\left|\frac{1}{V_{n,i}\left(t\right)}\sum_{\tau\in\mathcal{G}_{n,i}^{A}}r_{n,i}\left(\tau\right)-\mu_{n,i}\right|>\Delta\,|\,V_{\min}\geq v\right)\underset{\left(b\right)}{\leq}2NMe^{-\frac{\Delta^{2}}{2\sigma_{\max}^{2}+2b_{\max}\Delta}v}.\label{eq:26}
\end{multline}
 where (a) follows by taking the union bound over all players and
arms and (b) from using Bernstein's inequality (see \cite[Page 205]{1937}).
Since the exploration phase consists of uniform and independent arm
choices we have
\begin{equation}
\mathbb{P}\left(A_{n,i}\left(t\right)=1\right)=\frac{1}{M}\left(1-\frac{1}{M}\right)^{N-1}.\label{eq:27}
\end{equation}
Therefore
\begin{multline}
\mathbb{P}\left(V_{\min}<\frac{T_{e}\left(k\right)}{5M}\right)=\mathbb{P}\left(\bigcup_{i=1}^{M}\bigcup_{n=1}^{N}\left\{ V_{n,i}\left(t\right)<\frac{T_{e}\left(k\right)}{5M}\right\} \right)\underset{\left(a\right)}{\leq}NM\mathbb{P}\left(V_{1,1}\left(t\right)<\frac{T_{e}\left(k\right)}{5M}\right)\underset{\left(b\right)}{\leq}\\
NMe^{-2\frac{1}{M^{2}}\left(\left(1-\frac{1}{M}\right)^{N-1}-\frac{1}{5}\right)^{2}T_{e}\left(k\right)}\underset{\left(c\right)}{\leq}NMe^{-\frac{1}{18M^{2}}T_{e}\left(k\right)}\label{eq:28}
\end{multline}
where (a) follows from the union bound, (b) from Hoeffding's inequality
for Bernoulli random variables and (c) since $M\geq N$ and $\left(1-\frac{1}{M}\right)^{M-1}-\frac{1}{5}\geq e^{-1}-\frac{1}{5}>\frac{1}{6}$.
We conclude that
\begin{multline}
\mathbb{P}\left(E\right)=\sum_{v=0}^{T_{e}\left(k\right)}\mathbb{P}\left(E|V_{\min}=v\right)\mathbb{P}\left(V_{\min}=v\right)\leq\sum_{v=0}^{\left\lfloor \frac{T_{e}\left(k\right)}{5M}\right\rfloor }\mathbb{P}\left(V_{\min}=v\right)+\\
\sum_{v=\left\lfloor \frac{T_{e}\left(k\right)}{5M}\right\rfloor +1}^{T_{e}\left(k\right)}\mathbb{P}\left(E|V_{\min}=v\right)\mathbb{P}\left(V_{\min}=v\right)\leq\mathbb{P}\left(V_{\min}<\frac{T_{e}\left(k\right)}{5M}\right)+\mathbb{P}\left(E|\,\,V_{\min}\geq\frac{T_{e}\left(k\right)}{5M}\right)\underset{\left(a\right)}{\leq}\\
2NMe^{-\frac{\Delta^{2}T_{e}\left(k\right)}{M\left(10\sigma_{\max}^{2}+10b_{\max}\Delta\right)}}+NMe^{-\frac{T_{e}\left(k\right)}{18M^{2}}}\underset{\left(b\right)}{\leq}2NMe^{-\frac{\Delta^{2}c_{1}\left(\frac{k}{2}\right)^{\delta}}{M\left(20\sigma_{\max}^{2}+20b_{\max}\Delta\right)}k}+NMe^{-\frac{c_{1}\left(\frac{k}{2}\right)^{\delta}}{36M^{2}}k}\label{eq:29}
\end{multline}
where (a) follows from \eqref{eq:26} and \eqref{eq:28}, and (b)
from \eqref{eq:25}. By requiring $\Delta=\frac{\Delta_{\min}}{4}\leq\frac{J_{1}-J_{2}}{4N}$
we know from Lemma \ref{lem:Percision} that $\mathbb{P}\left(\boldsymbol{a}^{k*}\notin\mathcal{A}^{*}\right)\leq\mathbb{P}\left(E\right)$,
which together with \eqref{eq:29} establishes \eqref{eq:23}. Now
define $w$ as in \eqref{eq:22}. We obtain 
\begin{multline}
\mathbb{P}\left(\bigcup_{r=0}^{\left\lfloor \frac{k}{2}\right\rfloor }P_{e,k-r}\right)\underset{\left(a\right)}{\leq}2NM\sum_{r=0}^{\left\lfloor \frac{k}{2}\right\rfloor }e^{-wc_{1}\left(\frac{k-r}{2}\right)^{\delta}\left(k-r\right)}+NM\sum_{r=0}^{\left\lfloor \frac{k}{2}\right\rfloor }e^{-\frac{c_{1}}{36M^{2}}\left(\frac{k-r}{2}\right)^{\delta}\left(k-r\right)}\underset{\left(b\right)}{\leq}\\
2NMe^{-wc_{1}\left(\frac{k}{4}\right)^{\delta}k}\sum_{r=0}^{\left\lfloor \frac{k}{2}\right\rfloor }e^{wc_{1}\left(\frac{k}{4}\right)^{\delta}r}+NMe^{-\frac{c_{1}}{36M^{2}}\left(\frac{k}{4}\right)^{\delta}k}\sum_{r=0}^{\left\lfloor \frac{k}{2}\right\rfloor }e^{\frac{c_{1}}{36M^{2}}\left(\frac{k}{4}\right)^{\delta}r}\underset{\left(c\right)}{\leq}\\
2NMe^{-wc_{1}\left(\frac{k}{4}\right)^{\delta}k}\frac{e^{wc_{1}\left(\frac{k}{4}\right)^{\delta}\left(\frac{k}{2}+1\right)}-1}{e^{wc_{1}\left(\frac{k}{4}\right)^{\delta}}-1}+NMe^{-\frac{c_{1}}{36M^{2}}\left(\frac{k}{4}\right)^{\delta}k}\frac{e^{\frac{1}{36M^{2}}c_{1}\left(\frac{k}{4}\right)^{\delta}\left(\frac{k}{2}+1\right)}-1}{e^{\frac{1}{36M^{2}}c_{1}\left(\frac{k}{4}\right)^{\delta}}-1}\leq\\
\frac{2NM}{1-e^{-wc_{1}\left(\frac{k}{4}\right)^{\delta}}}e^{-\frac{w}{2}c_{1}\left(\frac{k}{4}\right)^{\delta}k}+\frac{NM}{1-e^{-\frac{1}{36M^{2}}c_{1}\left(\frac{k}{4}\right)^{\delta}}}e^{-\frac{1}{72M^{2}}c_{1}\left(\frac{k}{4}\right)^{\delta}k}\label{eq:30}
\end{multline}
where (a) follows by using the union bound on \eqref{eq:29} , (b)
follows since $e^{-wc_{1}\left(\frac{k-r}{2}\right)^{\delta}\left(k-r\right)}\leq e^{-wc_{1}\left(\frac{k}{4}\right)^{\delta}\left(k-r\right)}$
for $r\leq\left\lfloor \frac{k}{2}\right\rfloor $, and (c) is the
geometric sum formula.
\end{IEEEproof}

\section{Game of Thrones Dynamics Phase\label{sec:Game-of-Thrones}}

In this section, we analyze the game of thrones (GoT) dynamics between
players. These dynamics guarantee that the optimal state will be played
a significant amount of time, and only require the players to know
their own action and their received payoff on each turn. Note that
these dynamics assume deterministic utilities. We use the estimated
expected reward of each arm as the utility for this step, and zero
if a collision occurred. This means that players ignore the numerical
reward they receive by choosing the arm, as long as it is non-zero.
\begin{defn}
The game of thrones $G$ of epoch $k$ has the $N$ players of the
original multi-armed bandit game. Each player can choose from among
the $M$ arms, so $\mathcal{A}_{n}=\left\{ 1,...,M\right\} $ for
each $n$. The utility of player $n$ in the strategy profile $\boldsymbol{a}=\left(a_{1},...,a_{N}\right)$
is
\begin{equation}
u_{n}\left(\boldsymbol{a}\right)=\mu_{n,a_{n}}^{k}\eta_{a_{n}}\left(\boldsymbol{a}\right)\label{eq:31}
\end{equation}
where $\mu_{n,a_{n}}^{k}$ is the estimation of the expected reward
of arm $a_{n}$, from all the exploration phases that have ended,
up to epoch $k$. Also define $u_{n,\max}=\underset{\boldsymbol{a}}{\max}\,u_{n}\left(\boldsymbol{a}\right)$. 
\end{defn}
Our dynamics belong to the family introduced in \cite{Marden2014,Pradelski2012,Menon2013}.
These are very powerful dynamics that guarantee that the optimal strategy
profile (in terms of the sum of utilities) will be played a sufficiently
large portion of the turns. However, \cite{Marden2014,Pradelski2012,Menon2013}
all rely on the following structural property of the game, called
interdependence.
\begin{defn}
A game $G$ with finite action spaces $\mathcal{A}_{1},...,\mathcal{A}_{N}$
is interdependent if for every strategy profile $\boldsymbol{a}\in\mathcal{A}_{1}\times...\times\mathcal{A}_{N}$
and every set of players $J\subset N$, there exists a player $n\notin J$
and a choice of actions $\boldsymbol{a}_{J}'\in\prod_{m\in J}\mathcal{A}_{m}$
such that $u_{n}\left(\boldsymbol{a}_{J}',\boldsymbol{a}_{-J}\right)\neq u_{n}\left(\boldsymbol{a}_{J},\boldsymbol{a}_{-J}\right)$. 
\end{defn}
Our GoT is not interdependent. To see this, pick any strategy profile
$\boldsymbol{a}$ such that some players are in a collision while
others are not. Choose $J$ as the set of all players that are not
in a collision. All players outside this set are in a collision, and
there does not exist any colliding player such that the actions of
the non-colliding players can make her utility non-zero. 

Our GoT Dynamics, described in Subsection \ref{subsec:Game-of-Thrones-Dynamics},
modify \cite{Marden2014} such that interdependency is no longer needed.
In comparison to \cite{Marden2014}, our dynamics assign zero probability
that a player with $u_{n}=0$ (in a collision) will be content. Additionally,
we do not need to keep the benchmark utility as part of the state.
A player knows with probability 1 whether there was a collision, and
if there was not, she gets the same utility for the same arm. 

Unlike the exploration during the exploration phase, the exploration
of the GoT dynamics is meant to allow the players to reach the optimal
strategy profile. Since the GoT must have well-defined and time-invariant
utility functions, the reward samples from the GoT phase cannot be
used to change the estimation for the arm expectations already in
the current epoch. However, players can use these samples in the next
epoch and improve the accuracy of their estimation by doing so. While
it does not improve our regret analysis, it can enhance the algorithm's
performance in practice. 

The GoT dynamics (see Subsection 3.1) induce a Markov chain over the
state space $Z=\prod_{n=1}^{N}\left(\mathcal{A}_{n}\times\mathcal{M}\right)$,
where $\mathcal{M}=\left\{ C,D\right\} $. We denote the transition
matrix of this Markov chain by $P^{\varepsilon}$. We are interested
in the invariant distribution of $P^{\varepsilon}$, which, for small
$\varepsilon$, is concentrated only on the states with an optimal
sum of utilities. However, it is only guaranteed that the dynamics
will visit these optimal states very often for a small enough $\varepsilon$.
There could be multiple optimal states and the dynamics might fluctuate
between them. This could prevent players on distributedly agreeing
which optimal state to play. Fortunately, as shown in the following
lemma, in our case there is a unique optimal state with probability
one. This result arises from the continuous distribution of the rewards
that makes the distribution of the empirical estimation for the expectations
continuous as well. 
\begin{lem}
\label{lem:Uniqueness} The optimal solution to $\underset{\boldsymbol{a}}{\max}\sum_{n=1}^{N}u_{n}\left(\boldsymbol{a}\right)$
is unique with probability 1.
\end{lem}
\begin{IEEEproof}
First note that an optimal solution must not have any collisions,
otherwise it can be improved since $M\geq N$. Let $\left\{ \mu_{n,i}^{k}\right\} $
be the estimated reward expectations in epoch $k$. For two different
solutions $\tilde{\boldsymbol{a}}\neq\boldsymbol{a}^{*}$ to be optimal,
we must have $\sum_{n=1}^{N}\mu_{n,\tilde{a}_{n}}^{k}=\sum_{n=1}^{N}\mu_{n,a_{n}^{*}}^{k}$.
However, $\tilde{\boldsymbol{a}}$ and $\boldsymbol{a}^{\boldsymbol{*}}$
must differ in at least one assignment. Since the distributions of
the rewards $r_{n,a_{n}}$ are continuous, so are the distributions
of $\sum_{n=1}^{N}\mu_{n,a_{n}}^{k}$ (as a sum of the average of
the rewards). Hence $\mathbb{P}\left(\sum_{n=1}^{N}\mu_{n,\tilde{a}_{n}}^{k}=\sum_{n=1}^{N}\mu_{n,a_{n}^{*}}^{k}\right)=0$,
which is a contradiction, so the result follows. 
\end{IEEEproof}
Next we prove a lower bound on the probability for the unique optimal
state $z^{*}$ in the stationary distribution of the GoT dynamics.
This lower bound is a function of $\varepsilon$, and hence allows
for a choice of $\varepsilon$ that guarantees that $\pi_{z^{*}}>\frac{1}{2}$.
Note that the analysis in \cite{Marden2014} cannot be applied here
since our game is not interdependent. Moreover, our proof only requires
that $c>\sum_{n}u_{n,\max}-J_{1}$ where $J_{1}$ is the optimal objective.
This significantly improves the $c>N$ requirement in \cite{Marden2014},
and has a crucial effect on the mixing time of the GoT dynamics and
therefore on the convergence time to the optimal state. Note that
the analysis in \cite{Marden2014} assumed utilities that are in $\left[0,1\right]$
so $\sum_{n}u_{n,\max}\leq N$.

The following notion is often useful for the analysis of perturbed
Markov chains.
\begin{defn}
Let $Z$ be a finite set and let $z_{r}\in Z$. A graph consisting
of edges $z'\rightarrow z$ such that $z'\in Z\setminus\left\{ z_{r}\right\} $
and $z\in Z$ is called a $\left\{ z_{r}\right\} $-graph if it satisfies
the following conditions:
\begin{enumerate}
\item Every point $z'\in Z\setminus\left\{ z_{r}\right\} $ is the initial
point of exactly one edge. 
\item For any point $z'\in Z\setminus\left\{ z_{r}\right\} $ there exists
a sequence of edges leading from it to $z_{r}$.
\end{enumerate}
We denote by $G\left(z_{r}\right)$ the set of $\left\{ z_{r}\right\} $-graphs
and by $g$ a specific graph. A graph in $G\left(z_{r}\right)$ is
a tree rooted in $z_{r}$ such that from every $z'\neq z_{r}$ there
is a path to $z_{r}$. This follows since the two conditions above
imply that there are no closed cycles in a $\left\{ z_{r}\right\} $-graph. 
\end{defn}
The following lemma provides an explicit expression for the stationary
distribution of a Markov chain.
\begin{lem}[{\cite[Lemma 3.1, Chapter 6]{Freidlin1998}}]
\label{lem:Stationary}Consider a Markov chain with a set of states
$Z$ and transition probabilities $P_{z'z}$. Assume that every state
can be reached from any other state in a finite number of steps. Then
the stationary distribution of the chain is 
\begin{equation}
\pi\left(z\right)=\frac{Q\left(z\right)}{\sum_{z'\in Z}Q\left(z'\right)}\label{eq:32}
\end{equation}
where
\begin{equation}
Q\left(z\right)=\sum_{g\in G\left(z\right)}\prod_{\left(z'\rightarrow z\right)\in g}P_{z'z}.\label{eq:33}
\end{equation}
\end{lem}
Using the lemma above, we can prove the following lower bound on $\pi_{z^{*}}$
as a function of $\varepsilon$. 
\begin{thm}
\label{thm:StationaryBound}Let $\varepsilon>0$ and let $\pi$ be
the stationary distribution of $Z$. Let $\boldsymbol{a}^{k*}=\arg\underset{\boldsymbol{a}}{\max}\sum_{n=1}^{N}u_{n}\left(\boldsymbol{a}\right)$
and let the optimal state be $z^{*}=$$\left[\boldsymbol{a}^{k*},C^{N}\right]$.
Let $J_{1}=\sum_{n=1}^{N}u_{n}\left(\boldsymbol{a}^{k*}\right)$.
If $c>\sum_{n}u_{n,\max}-J_{1}$ then for any $0<\eta<\frac{1}{2}$
there exists a small enough $\varepsilon$ such that $\pi_{z^{*}}>\frac{1}{2\left(1-\eta\right)}$.
\end{thm}
\begin{IEEEproof}
See Appendix A. 
\end{IEEEproof}
We need that $\pi_{z^{*}}>\frac{1}{2}$ in order for the GoT phase
to succeed. The main role of the theorem above is to show that a small
enough $\varepsilon$, such that $\pi_{z^{*}}>\frac{1}{2}$, exists.
However, the proof of Theorem \ref{thm:StationaryBound} also tells
which $\varepsilon$ values are small enough to guarantee that $\pi_{z^{*}}>\frac{1}{2}$,
as a function of the problem parameters $N,M$ and the matrix of utilities
$\left\{ u_{n}\right\} $. The designer that has to choose $\varepsilon$
does not typically know the parameters of the problem. Nevertheless,
even having a coarse bound on these parameters is enough since we
only need to choose a small enough $\varepsilon$ instead of tuning
$\varepsilon$ as a function of these parameters. In practice, the
designer is likely to have either uncertainty bounds for the problem
parameters or a random model that is either known or can be simulated.
Using this random model, the designer can always take $\varepsilon$
small enough to make the probability that $\pi_{z^{*}}>\frac{1}{2}$
arbitrarily close to one. 

Next we prove a probabilistic lower bound on the number of times the
optimal state has been played during the $k$-th GoT phase. 
\begin{lem}
\label{lem:Convergence Lemma} Let $\boldsymbol{a}^{k*}=\arg\underset{\boldsymbol{a}}{\max}\sum_{n=1}^{N}u_{n}\left(\boldsymbol{a}\right)$
and let the optimal state be $z^{*}=$$\left[\boldsymbol{a}^{k*},C^{N}\right]$.
Denote the stationary distribution of $Z$ by $\pi$. Let $\mathcal{G}_{k}$
be the set of turns of the $k$-th GoT phase. Then for any $0<\eta<1$
we have, for a sufficiently large $k$, that
\begin{equation}
P_{g}\triangleq\mathbb{P}\left(\sum_{t\in\mathcal{G}_{k}}I\left(z\left(t\right)=z^{*}\right)\leq\left(1-\eta\right)\pi_{z^{*}}\left\lceil c_{2}k^{\delta}\right\rceil \right)\leq\underbrace{B_{0}\left\Vert \varphi_{k}\right\Vert _{\pi}}_{A_{k}}e^{-\frac{\eta^{2}\pi_{z^{*}}c_{2}k^{\delta}}{72T_{m}\left(\frac{1}{8}\right)}}\label{eq:34}
\end{equation}
where $I$ is the indicator function, $B_{0}$ is a constant independent
of $\eta$ and $\pi_{z^{*}}$, $\varphi_{k}$ is the probability distribution
of the state played in the $k-\left\lfloor \frac{k}{2}\right\rfloor -1$-th
exploitation phase, and $\left\Vert \varphi_{k}\right\Vert _{\pi}\triangleq\sqrt{\sum_{i=1}^{\left|Z\right|}\frac{\varphi_{k,i}^{2}}{\pi_{i}}}$.
\end{lem}
\begin{IEEEproof}
The events $I\left(z\left(t\right)=z^{*}\right)$ are not independent
but rather form a Markov chain. Hence, Markovian concentration inequalities
are required. The result follows by a simple application of the concentration
bound in \cite[Theorem 3]{Chung2012}. We use $f\left(z\right)=I\left(z=z^{*}\right)$,
which counts the number of visits to the optimal state. We denote
by $T_{m}\left(\frac{1}{8}\right)$ the mixing time of $Z$ with an
accuracy of $\frac{1}{8}$. Our initial state is the state $\hat{z}_{k}$
played in the $k-\left\lfloor \frac{k}{2}\right\rfloor -1$ exploitation
phase.
\end{IEEEproof}
The next lemma concludes this section by providing an upper bound
for the probability that the $k$-th exploitation phase failed due
to the last $\left\lfloor \frac{k}{2}\right\rfloor +1$ GoT phases
(including the $k$-th GoT phase).
\begin{lem}[GoT Error Probability]
\label{lem:ConnectingGoTs} Let $\boldsymbol{a}^{k*}=\arg\underset{\boldsymbol{a}}{\max}\sum_{n=1}^{N}u_{n}\left(\boldsymbol{a}\right)$.
Let the optimal state of the $k$-th GoT phase be $z^{k*}=$$\left[\boldsymbol{a}^{k*},C^{N}\right]$.
Let $\mathcal{G}_{k}$ be the set of turns of the $k$-th GoT phase.
Define the number of times the optimal state has been played in the
last $\left\lfloor \frac{k}{2}\right\rfloor +1$ GoT phases combined
by 
\begin{equation}
F_{k}\left(z^{*}\right)\triangleq\sum_{i=k-\left\lfloor \frac{k}{2}\right\rfloor }^{k}\sum_{t\in\mathcal{G}_{i}}I\left(z\left(t\right)=z^{i*}\right).\label{eq:37}
\end{equation}
Let $\pi_{z^{*}}=\underset{k-\left\lfloor \frac{k}{2}\right\rfloor \leq i\leq k}{\min}\pi_{z^{i*}}$
and $T_{m}\left(\frac{1}{8}\right)=\underset{k-\left\lfloor \frac{k}{2}\right\rfloor \leq i\leq k}{\max}T_{m,i}\left(\frac{1}{8}\right)$.
If $\pi_{z^{*}}>\frac{1}{2\left(1-\eta\right)}$ for some $0<\eta<\frac{1}{2}$
then for a sufficiently large $k$ 
\begin{equation}
P_{c,k}\triangleq\mathbb{P}\left(F_{k}\left(z^{*}\right)\leq\frac{1}{2}\sum_{i=k-\left\lfloor \frac{k}{2}\right\rfloor }^{k}c_{2}i^{\delta}\right)\leq\left(C_{0}e^{-\frac{c_{2}\eta^{2}}{144T_{m}\left(\frac{1}{8}\right)}\left(\pi_{z^{*}}-\frac{1}{2\left(1-\eta\right)}\right)\left(\frac{k}{2}\right)^{\delta}}\right)^{k}.\label{eq:38}
\end{equation}
where $C_{0}$ is a constant that is independent of $k$, $\pi_{z^{*}}$
and $\eta$. 
\end{lem}
\begin{IEEEproof}
Define the total length of the last $\left\lfloor \frac{k}{2}\right\rfloor +1$
GoT phases by $L_{k}=c_{2}\sum_{i=k-\left\lfloor \frac{k}{2}\right\rfloor }^{k}i^{\delta}$.
Define the independent (given $\left\{ z^{i}\left(0\right)\right\} _{i}$)
random variables $S_{i}=\sum_{t\in\mathcal{G}_{i}}I\left(z\left(t\right)=z^{i*}\right)$
for each $i$. For any $t>0$, Chernoff bound on \eqref{eq:37} yields
\begin{equation}
\mathbb{P}\left(\sum_{i=k-\left\lfloor \frac{k}{2}\right\rfloor }^{k}S_{i}\leq\frac{L_{k}}{2}\,|\,\left\{ z^{i}\left(0\right)\right\} _{i=k-\left\lfloor \frac{k}{2}\right\rfloor }^{k}\right)\leq e^{t\frac{L_{k}}{2}}\prod_{i=k-\left\lfloor \frac{k}{2}\right\rfloor }^{k}\mathbb{E}\left\{ e^{-tS_{i}}\,|\,z^{i}\left(0\right)\right\} \label{eq:39}
\end{equation}
where $z^{i}\left(0\right)$ is the initial state at the beginning
of the $i$-th GoT phase. For $t=\frac{\frac{\eta^{2}}{1-\eta}}{72T_{m}\left(\frac{1}{8}\right)}$
we have the following bound 
\begin{equation}
\mathbb{E}\left\{ e^{-tS_{i}}\,|\,z^{i}\left(0\right)\right\} \underset{\left(a\right)}{\leq}P_{g}\cdot1+\left(1-P_{g}\right)\cdot e^{-t\left(1-\eta\right)\pi_{z^{*}}c_{2}i^{\delta}}\underset{\left(b\right)}{\leq}\left(1+A_{i}\right)e^{-\frac{\eta^{2}\pi_{z^{*}}c_{2}i^{\delta}}{72T_{m}\left(\frac{1}{8}\right)}}\label{eq:40}
\end{equation}
where (a) is since $e^{-tS_{i}}\leq1$ and (b) follows from Lemma
\ref{lem:Convergence Lemma} on $P_{g}$. Plugging \eqref{eq:40}
back into \eqref{eq:39} and using the law of total expectation over
$\left\{ z^{i}\left(0\right)\right\} _{i=k-\left\lfloor \frac{k}{2}\right\rfloor }^{k}$
gives
\begin{multline}
\mathbb{P}\left(\sum_{i=k-\left\lfloor \frac{k}{2}\right\rfloor }^{k}S_{i}\leq\frac{L_{k}}{2}\right)\underset{\left(a\right)}{\leq}\left[\prod_{i=k-\left\lfloor \frac{k}{2}\right\rfloor }^{k}\left(1+\mathbb{E}^{z^{i}\left(0\right)}\left\{ A_{i}\right\} \right)\right]e^{\frac{\frac{\eta^{2}}{1-\eta}}{72T_{m}\left(\frac{1}{8}\right)}\frac{L_{k}}{2}}e^{-\frac{\eta^{2}\pi_{z^{*}}}{72T_{m}\left(\frac{1}{8}\right)}L_{k}}\underset{\left(b\right)}{\leq}\\
C_{0}^{k}e^{-\frac{\eta^{2}}{72T_{m}\left(\frac{1}{8}\right)}\left(\pi_{z^{*}}-\frac{1}{2\left(1-\eta\right)}\right)L_{k}}\underset{\left(c\right)}{\leq}\left(C_{0}e^{-\frac{c_{2}\eta^{2}}{144T_{m}\left(\frac{1}{8}\right)}\left(\pi_{z^{*}}-\frac{1}{2\left(1-\eta\right)}\right)\left(\frac{k}{2}\right)^{\delta}}\right)^{k}\label{eq:41}
\end{multline}
where (a) uses $\prod_{i=k-\left\lfloor \frac{k}{2}\right\rfloor }^{k}e^{-ac_{2}i^{\delta}}=e^{-a\sum_{i=k-\left\lfloor \frac{k}{2}\right\rfloor }^{k}c_{2}i^{\delta}}$for
any $a$. In (b), note that for every $z^{i}\left(0\right)$ we have
$A_{i}=c\left\Vert \varphi_{i}\right\Vert _{\pi}$ for some constant
$c$, so $\mathbb{E}^{z^{i}\left(0\right)}\left\{ A_{i}\right\} $
is bounded for any $i$ and the above bound vanishes with $k$. However,
by choosing $z^{i}\left(0\right)$ to be the state played in the $i-\left\lfloor \frac{i}{2}\right\rfloor -1$
exploitation phase, the same bound guarantees that $\mathbb{P}\left(z^{i}\left(0\right)=z^{i*}\right)\rightarrow1$
as $i\rightarrow\infty$, so $\mathbb{E}\left\{ \left\Vert \varphi_{i}\right\Vert _{\pi}\right\} \rightarrow\frac{1}{\sqrt{\pi_{z^{i*}}}}<\sqrt{2}.$
Inequality (c) uses $L_{k}=\sum_{i=k-\left\lfloor \frac{k}{2}\right\rfloor }^{k}c_{2}i^{\delta}\geq\frac{k}{2}\left(\frac{k}{2}\right)^{\delta}$.
Note that the counting in \eqref{eq:37} is on the same optimal state
$z^{i*}=z^{*}$ for $k-\left\lfloor \frac{k}{2}\right\rfloor \leq i\leq k$
only if the previous $\left\lfloor \frac{k}{2}\right\rfloor +1$ exploration
phases succeeded. 
\end{IEEEproof}

\section{Markovian Rewards\label{sec:Markovian-Rewards}}

In this section, we generalize our main result to the case of (rested)
Markovian rewards. This generalization is valid for the same GoT algorithm
(Algorithm \ref{alg:Game-of-Thrones}) without any modifications.
In this section we assume that each player can observe her collision
indicator in addition to her reward. The reason for this additional
assumption is that with discrete rewards, the collision indicator
cannot be deduced from the rewards with probability 1. Nevertheless,
knowing whether other players chose the same arm is a very modest
requirement compared to assuming that players can observe the actions
of other players. 

We use the standard model for Markovian bandits (see for example \cite{Tekin2012}).
As in the i.i.d. case, we assume that the reward processes (now Markov
chains) of players are independent. 
\begin{defn}
\label{Def: Rested Markovian}Let $V_{n,i}$ be the number of visits
without collision of player $n$ to arm $i$. The sequence of rewards
$\left\{ r_{n,i}\left(V_{n,i}\right)\right\} $ of arm $i$ for player
$n$ is an ergodic Markov chain such that:
\begin{enumerate}
\item $r_{n,i}\left(V_{n,i}\right)$ has a finite state space $\mathcal{R}_{n,i}$
for each $n,i$, consisting of positive numbers. 
\item The transition matrix of $r_{n,i}\left(V_{n,i}\right)$ is $P_{n,i}$
and the stationary distribution is $\pi^{n,i}$.
\item The Markov chains $\left\{ r_{n,i}\left(V_{n,i}\right)\right\} $
are independent for different $n$ or different $i$.
\end{enumerate}
\end{defn}
The total regret is now defined as 
\begin{defn}
The expectation of arm $i$ for player $n$ is defined as:
\begin{equation}
\mu_{n,i}=\sum_{r\in\mathcal{R}_{n,i}}r\pi^{n,i}\left(r\right).\label{eq:42}
\end{equation}
Define $g_{n}\left(\boldsymbol{a}\right)=\mu_{n,a_{n}}\eta_{a_{n}}\left(\boldsymbol{a}\right)$
as the expected utility of player $n$ in strategy profile $\boldsymbol{a}$.
The total regret is defined as the random variable
\begin{equation}
R=\sum_{t=1}^{T}\sum_{n=1}^{N}g_{n}\left(\boldsymbol{a}^{*}\right)-\sum_{t=1}^{T}\sum_{n=1}^{N}r_{n,a_{n}\left(t\right)}\left(V_{n,i}\left(t\right)\right)\eta_{a_{n}\left(t\right)}\left(\boldsymbol{a}\left(t\right)\right)\label{eq:43}
\end{equation}
where
\begin{equation}
\boldsymbol{a}^{*}\in\arg\underset{\boldsymbol{a}}{\max}\sum_{n=1}^{N}g_{n}\left(\boldsymbol{a}\right).\label{eq:44}
\end{equation}
The expected total regret $\bar{R}\triangleq\mathbb{E}\left\{ R\right\} $
is the average of \eqref{eq:43} over the randomness of the rewards
$\left\{ r_{n,i}\left(V_{n,i}\right)\right\} $, that dictate the
random actions $\left\{ a_{n}\left(t\right)\right\} $.
\end{defn}
The division into epochs and phases makes the generalization of our
result to other reward models convenient. Only the exploration phase
requires a different analysis. This is formalized in the following
theorem. As before, an explicit requirement on $\varepsilon$ is given
in the proof of Theorem \ref{thm:StationaryBound}. 
\begin{thm}[Generalization to Markovian Rewards]
Assume that the rewards $\left\{ r_{n,i}\left(V_{n,i}\right)\right\} $
are Markovian as in Definition \ref{Def: Rested Markovian}. Assume
that for every $n,i$, the stationary distribution of $r_{n,i}\left(V_{n,i}\right)$,
denoted $\pi^{n,i}$, is generated at random using a continuous distribution
on the $M$ dimensional simplex. Let the game have a finite horizon
$T$, unknown to the players. Let each player play according to Algorithm
\ref{alg:Game-of-Thrones}, with any integers $c_{1},c_{2},c_{3}>0$
and $\delta>0$. Then there exists a small enough $\varepsilon$ such
that for large enough $T$, the expected total regret is bounded by
\begin{equation}
\bar{R}\leq4\left(\underset{n,i}{\max}\mu_{n,i}\right)\left(c_{1}+c_{2}\right)N\log_{2}^{1+\delta}\left(\frac{T}{c_{3}}+2\right)=O\left(\log^{1+\delta}T\right).\label{eq:45}
\end{equation}
\end{thm}
\begin{IEEEproof}
First note that an optimal solution must not have any collisions,
otherwise it can be improved since $M\geq N$ and the expected rewards
are positive. Since the stationary distributions are continuous, the
probability for any specific value of the expected rewards $\left\{ \mu_{n,i}\right\} $
defined in \eqref{eq:42}, or any specific value of $\sum_{n=1}^{N}\mu_{n,\tilde{a}_{n}}$,
is zero. For two different solutions $\tilde{\boldsymbol{a}}\neq\boldsymbol{a}^{*}$
to be optimal, they must have $\sum_{n=1}^{N}\mu_{n,\tilde{a}_{n}}=\sum_{n=1}^{N}\mu_{n,a_{n}^{*}}$.
However, they must differ in at least one assignment. Hence $\mathbb{P}\left(\sum_{n=1}^{N}\mu_{n,\tilde{a}_{n}}=\sum_{n=1}^{N}\mu_{n,a_{n}^{*}}\right)=0$,
so there is a unique solution to \eqref{eq:44} with probability 1.
If the exploration phase of the $k$-th epoch did not fail, then the
solution to $\underset{a}{\max}\sum_{n=1}^{N}\mu_{n,a_{n}}^{k}$ is
identical to this unique solution. \\
The rest of the proof replaces the exploration error bound in Lemma
\ref{lem: exploration}. The estimated expected rewards $\hat{\mu}_{n,i}$
are now defined with respect to the stationary distribution of the
reward chain. However, there is still a single estimated value for
each player and arm and the role and analysis of the GoT phase remain
the same. Hence, we need to show that the exploration phase results
in an accurate enough estimation of $\mu_{n,i}$.\\
As in \eqref{eq:25}, after the $k$-th exploration phase, the number
of samples that are used for estimating the expected rewards is $T_{e}\left(k\right)\geq c_{1}\left(\frac{k}{2}\right)^{\delta+1}$.
Let $T_{n,i}\left(\frac{1}{8}\right)$ be the mixing time of $r_{n,i}\left(t\right)$
with an accuracy of $\frac{1}{8}$. We define for the initial distribution
$\varphi$ on $\mathcal{R}_{n,i}$ 
\begin{equation}
\left\Vert \varphi\right\Vert _{\pi_{n,i}}\triangleq\sqrt{\sum_{j=1}^{\left|\mathcal{R}_{n,i}\right|}\frac{\varphi_{j}^{2}}{\pi_{n,i}\left(j\right)}}.\label{eq:46}
\end{equation}
Define $A_{n,i}\left(t\right)$ as the indicator that is equal to
one if only player $n$ chose arm $i$ at time $t$. Let $\mathcal{G}_{n,i}^{A}$
be the set of times for which $A_{n,i}\left(t\right)=1$. Define $V_{n,i}\left(t\right)\triangleq\sum_{\tau\in\mathcal{G}_{n,i}^{A}}A_{n,i}\left(\tau\right)$,
which is the number of visits of player $n$ to arm $i$ with no collision,
up to time $t$ and define $V_{\min}=\underset{n,i}{\min}V_{n,i}\left(t\right)$.
Define $E$ as the event in which there exists a player $n$ that
has an estimate of some arm $i$ with an accuracy worse than $\Delta$.
We have
\begin{multline}
\mathbb{P}\left(E|V_{\min}\geq v\right)=\mathbb{P}\left(\bigcup_{i=1}^{M}\bigcup_{n=1}^{N}\left\{ \left|\frac{1}{V_{n,i}\left(t\right)}\sum_{\tau\in\mathcal{G}_{n,i}^{A}}r_{n,i}\left(\tau\right)-\mu_{n,i}\right|\geq\Delta\,|\,V_{\min}\geq v\right\} \right)\underset{\left(a\right)}{\leq}\\
NM\underset{n,i}{\max}\left(\mathbb{P}\left(\sum_{\tau\in\mathcal{G}_{n,i}^{A}}r_{n,i}\left(\tau\right)\geq\left(1+\frac{\Delta}{\mu_{n,i}}\right)\mu_{n,i}v\right)+\mathbb{P}\left(\sum_{\tau\in\mathcal{G}_{n,i}^{A}}r_{n,i}\left(\tau\right)\leq\left(1-\frac{\Delta}{\mu_{n,i}}\right)\mu_{n,i}v\right)\right)\underset{\left(b\right)}{\leq}\\
2NMc\max_{n,i}\left(\left\Vert \varphi\right\Vert _{\pi_{n,i}}\right)e^{-\underset{n,i}{\min}\frac{\Delta\min\left\{ \frac{\Delta}{\mu_{n,i}},1\right\} }{72T_{n,i}\left(\frac{1}{8}\right)}v}\label{eq:47}
\end{multline}
where (a) follows by taking the union bound over all players and arms
and (b) from using the bound in \cite{Chung2012} with $\eta=\frac{\Delta}{\mu_{n,i}}$,
where $c$ is a constant independent of $\eta$ and $\pi_{z^{*}}$.
We conclude that
\begin{multline}
\mathbb{P}\left(E\right)=\sum_{v=0}^{T_{e}\left(k\right)}\mathbb{P}\left(E|V_{\min}=v\right)\mathbb{P}\left(V_{\min}=v\right)\leq\\
\sum_{v=0}^{\left\lfloor \frac{T_{e}\left(k\right)}{5M}\right\rfloor }\mathbb{P}\left(V_{\min}=v\right)+\sum_{v=\left\lfloor \frac{T_{e}\left(k\right)}{5M}\right\rfloor +1}^{T_{e}\left(k\right)}\mathbb{P}\left(E|V_{\min}=v\right)\mathbb{P}\left(V_{\min}=v\right)\leq\mathbb{P}\left(V_{\min}<\frac{T_{e}\left(k\right)}{5M}\right)+\\
\mathbb{P}\left(E|\,\,V_{\min}\geq\frac{T_{e}\left(k\right)}{5M}\right)\underset{\left(a\right)}{\leq}2NMc\max_{n,i}\left(\left\Vert \varphi\right\Vert _{\pi_{n,i}}\right)e^{-\underset{n,i}{\min}\frac{\Delta\min\left\{ \frac{\Delta}{\mu_{n,i}},1\right\} c_{1}\left(\frac{k}{2}\right)^{\delta}}{720T_{n,i}\left(\frac{1}{8}\right)M}k}+NMe^{-\frac{c_{1}\left(\frac{k}{2}\right)^{\delta}}{36M^{2}}k}.\label{eq:48}
\end{multline}
where (a) follows from \eqref{eq:47} and \eqref{eq:28}, that still
holds since arm choices are independent and uniform in the exploration
phase. Finally, by requiring $\Delta=\frac{\Delta_{\min}}{4}=\frac{1}{4}\underset{n}{\min}\underset{\boldsymbol{a}_{1},\boldsymbol{a}_{2}\,|\,g_{n}\left(\boldsymbol{a}_{1}\right)\neq g_{n}\left(\boldsymbol{a}_{2}\right)}{\min}\left|g_{n}\left(\boldsymbol{a}_{1}\right)-g_{n}\left(\boldsymbol{a}_{2}\right)\right|\leq\frac{J_{1}-J_{2}}{4N}$
we know from Lemma \ref{lem:Percision} that $\mathbb{P}\left(\boldsymbol{a}^{k*}\notin\mathcal{A}^{*}\right)\leq\mathbb{P}\left(E\right)$,
which together with \eqref{eq:48} establishes the same bound (up
to a constant factor) as in \eqref{eq:23} but with $\tilde{w}=\frac{\Delta_{\min}^{2}}{11520M\underset{n,i}{\max}\left[\mu_{n,i}T_{n,i}\left(\frac{1}{8}\right)\right]}$
replacing $w$. Using this new bound, the proof of Theorem \ref{thm:Main}
remains the same using $\tilde{w}$ instead of $w$.
\end{IEEEproof}

\section{Simulation Results\label{sec:Simulation-Results}}

In this section, we illustrate the behavior of Algorithm \ref{alg:Game-of-Thrones}
using numerical simulations. We use $\delta=0$ since it yields good
results in practice. We conjecture that the bound \eqref{eq:34} is
not tight for our particular Markov chain and indicator function,
since it applies to all Markov chains with the same mixing time and
all functions on the states. This explains why modest choices of $c_{2}$
are large enough to satisfy \eqref{eq:17} even for $\delta=0$, making
the $k^{\delta}$ factor in the exponent unnecessary in practice.
In order not to waste the initial phases, their lengths should be
chosen so that the exploitation phase already occupies most of the
turns in early epochs, while allowing for a considerable GoT phase.
In the implementation of our algorithm we also use the GoT and exploration
phases for estimating the expected rewards, which improves the estimation
significantly in practice, making the GoT phase the main issue. We
chose the parameter $c=\frac{\log\left(\frac{2}{c_{2}N}\right)}{\log\varepsilon}\approx1.4$
(for $\varepsilon=0.01$), in order for the escape probability from
a content state to be $\varepsilon^{c}=\frac{2}{c_{2}N}$, with an
expectation of approximately 2 escapes in each GoT phase. Note that
this choice is a significant relaxation from $c=N$ (the lowest possible
value in \cite{Marden2014}), and has a dramatic positive effect on
the mixing time of the GoT Dynamics and the convergence time. The
rewards are generated as $r_{n,i}\left(t\right)=\mu_{n,i}+z_{n,i}\left(t\right)$
where $\left\{ \mu_{n,i}\right\} $ are taken from a matrix $U$ and
$\left\{ z_{n,i}\left(t\right)\right\} $ are independent Gaussian
variables with zero mean and a variance of $\sigma^{2}=0.05$ for
each $n,i$.

\begin{figure*}[tbh]
\begin{minipage}[t]{0.5\columnwidth}%
~~~~~\includegraphics[width=9cm,height=5cm,keepaspectratio]{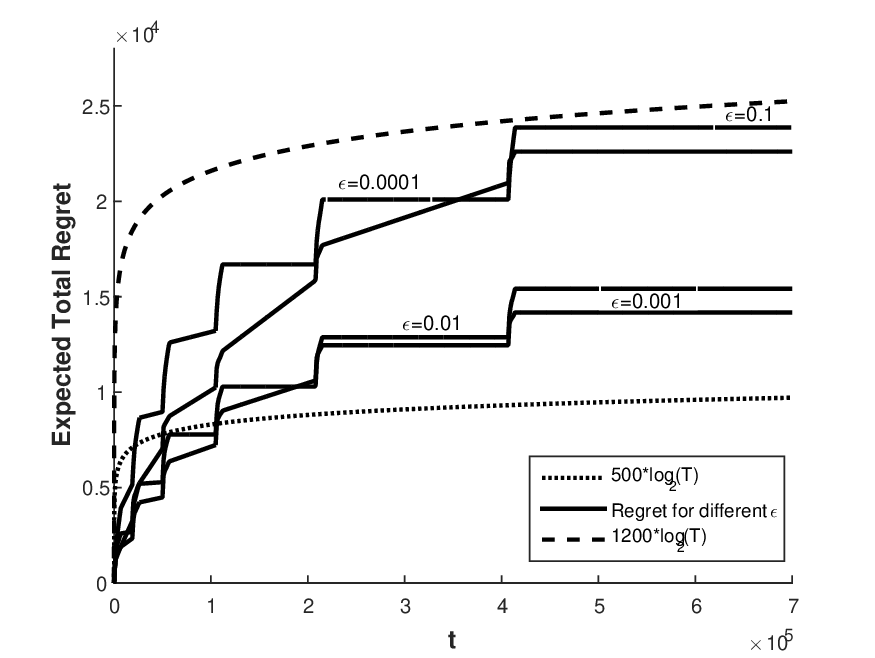}\subcaption{}\label{subfiga}%
\end{minipage}%
\begin{minipage}[t]{0.5\columnwidth}%
~\includegraphics[width=9cm,height=5cm,keepaspectratio]{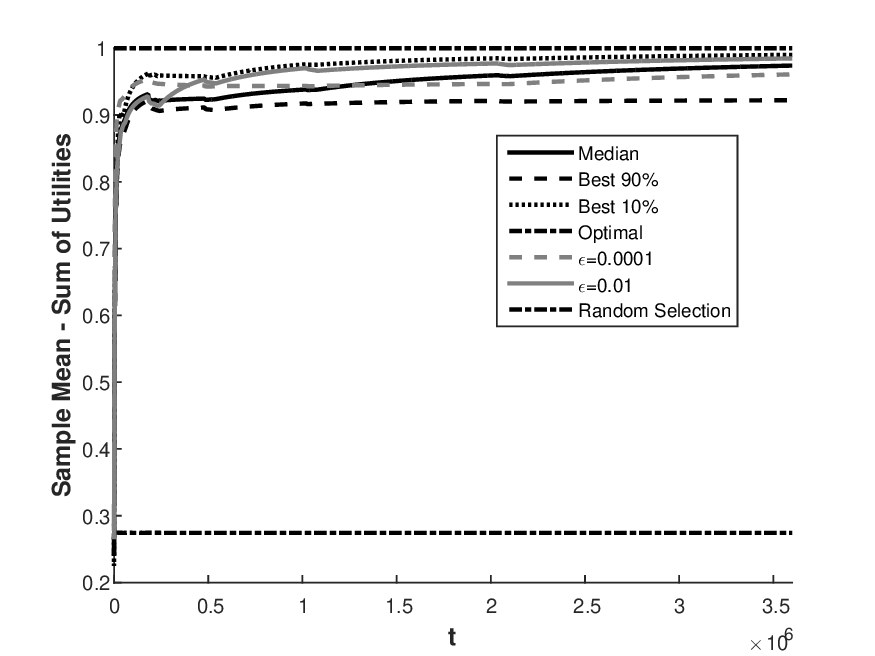}\subcaption{}\label{subfigb}%
\end{minipage}

\caption{ \label{fig:Total Regret}In (a): Total Regret for $N=M=3$ as a function
of time. In (b): Sample mean of the sum of utilities for $N=M=5$
as a function of time. Both figures are averaged over 100 simulations.}
\end{figure*}

First we demonstrate our theoretical result in the following scenario
with $N=M=3$:

{\scriptsize{}
\[
U=\left(\begin{array}{ccc}
0.1 & 0.05 & 0.9\\
0.1 & 0.25 & 0.3\\
0.4 & 0.2 & 0.8
\end{array}\right)
\]
}{\scriptsize\par}

for which the optimal allocation is $a_{1}=3,a_{2}=2,a_{1}=1$. Here
we used $c_{1}=500,c_{2}=c_{3}=6000$. The expected total regret as
a function of time, averaged over 100 realizations of the algorithm,
is depicted in part (a) of Fig. \ref{fig:Total Regret}. After the
second epoch, all epochs lead to the optimal solution in their exploitation
phase. It can be seen that the regret increases like $O\left(\log_{2}T\right)$,
and more specifically, is between $400\log_{2}T$ and $700\log_{2}T$.
This demonstrates the theoretical result of Theorem \ref{thm:Main}.
Furthermore, we see that the regret behaves very similarly across
4 different orders of magnitude of $\varepsilon$. This suggests that
choosing a small enough $\varepsilon$ is easy in practice.

The total regret compares the sum of utilities to the ideal one that
could have been achieved in a centralized scenario. With no communication
between players and with a matrix of expected rewards, the gap from
this ideal naturally increases. In this scenario, converging to the
exact optimal solution might take a long time, even for the (unknown)
optimal algorithm. A nice property of our algorithm that makes it
appealing in practice is that the GoT dynamics are not specifically
oriented to converging to the optimal solution, but they probabilistically
prefer states with a higher sum of utilities. This is simply because
these states have incoming paths with high probabilities (i.e., ``low
resistance''). To demonstrate this property, we generated 100 independent
realizations of $U$ with elements that were chosen uniformly at random
on $\left[0.05,0.95\right]$. In part (b) of Fig. \ref{fig:Total Regret},
we present the sample mean of the accumulated sum of utilities $\sum_{n=1}^{N}\frac{1}{t}\sum_{\tau=1}^{t}u_{n}\left(\boldsymbol{a}\left(\tau\right)\right)$
as a function of time $t$ and averaged over 100 experiments for $M=N=5$.
Here we used $c_{1}=500,c_{2}=c_{3}=60000$ and $\varepsilon=0.001$.
The performance was normalized by the optimal solution to the assignment
problem (for each $U$) and compared to the performance of a random
choice of arms. Clearly the sum of utilities becomes close to optimal
(more than 90\%) fast, with only a small variation between different
realizations of $U$. Additionally, our algorithm behaved very similarly
for a wide range of $\varepsilon$ values (3 orders of magnitude).
This supports the intuition that there is no threshold phenomenon
on $\varepsilon$ (becoming ``small enough''), since the dynamics
prefer states with higher sum of utilities for all $\varepsilon<1$.
Furthermore, it indicates that there is no penalty for choosing smaller
$\varepsilon$ than necessary just to have a safety margin. 

\section{Conclusion and Open Questions}

In this paper, we considered a multi-player multi-armed bandit game
where players distributedly learn how to allocate the arms (i.e.,
resources) between them. In contrast to all other multi-player bandit
problems, we both allow for different expected rewards between users
\textbf{and} assume users only know their actions and rewards. We
proposed a novel fully distributed algorithm that achieves an expected
total regret of near-$O\left(\log T\right)$, when the horizon $T$
is unknown by the players. 

Our simulations suggest that tuning the parameters for our algorithm
is a relatively easy task in practice. The designer can do so by simulating
a random model for the unknown environment and varying the parameters,
knowing that only very slack accuracy is needed for the tuning. 

In the single-player case, if the algorithm knows enough about the
problem parameters, it is possible to even achieve a bounded regret
(see \cite{lai1984asymptotically}). It is an interesting question
whether a bounded regret can be achieved in the multi-player scenario
with enough knowledge of the problem parameters. 

Analytically, a single epoch in our algorithm converges to the optimal
solution of the assignment problem with high probability. This might
be valuable even outside the context of bandit learning algorithms,
since it is the first fully distributed algorithm that solves the
assignment problem exactly. Our dynamics may take a long time to converge
in terms of $N$. While this does not effect the regret for large
$T$, it might be significant if our algorithm were to be used for
this purpose. Our game is not a general one but has a structure that
allowed us to modify the dynamics in \cite{Marden2014} such that
the interdependence assumption can be dropped. We conjecture that
this structure can also be exploited to accelerate the converge rate
of this kind of dynamics. Studying the converge time of perhaps further
refined dynamics is another promising future research direction.

\section{Appendix A - Proof of Theorem \ref{thm:StationaryBound}}

In this section we prove Theorem \ref{thm:StationaryBound}, which
bounds from below the probability of $z^{*}$ in the stationary distribution
of $Z$. From Lemma \ref{lem:Stationary} we know that 
\begin{equation}
\pi_{z^{*}}=\frac{1}{1+\underbrace{\frac{\sum_{z_{r}\in Z,z_{r}\neq z^{*}}\sum_{g\in G\left(z_{r}\right)}\prod_{\left(z'\rightarrow z\right)\in g}P_{z'z}}{\sum_{g\in G\left(z^{*}\right)}\prod_{\left(z'\rightarrow z\right)\in g}P_{z'z}}}_{A}}.\label{eq:62}
\end{equation}
In Subsection \ref{subsec:The-Maximal-Probability}, we bound the
denominator of A from below by identifying the maximal term in the
sum (the trees in $G\left(z^{*}\right)$ with the maximal probability).
In Subsection \ref{subsec:rest of the trees} we upper bound the numerator
of A. 

\subsection{The Maximal Probability Tree\label{subsec:The-Maximal-Probability}}

For each $z_{r}$, the dominant term in \eqref{eq:33} is
\begin{equation}
\underset{g\in G\left(z_{r}\right)}{\max}\prod_{\left(z'\rightarrow z\right)\in g}P_{z'z}.\label{eq:49}
\end{equation}
Define $S_{z}$ as the personal state vector in $z$. We decompose
the transition probabilities as follows
\begin{equation}
P_{z'z}=\mathbb{P}\left(z\,|\,z'\right)=\mathbb{P}\left(S_{z},\bar{a}_{z}\,|\,S_{z'},\bar{a}_{z'}\right)=\mathbb{P}\left(\bar{a}_{z}\,|\,S_{z'},\bar{a}_{z'}\right)\mathbb{P}\left(S_{z}\,|\,S_{z'},\bar{a}_{z'},\bar{a}_{z}\right).\label{eq:50}
\end{equation}
So the objective becomes
\begin{equation}
\prod_{\left(z'\rightarrow z\right)\in g}P_{z'z}=\prod_{\left(z'\rightarrow z\right)\in g}\mathbb{P}\left(\bar{a}_{z}\,|\,S_{z'},\bar{a}_{z'}\right)\prod_{\left(z'\rightarrow z\right)\in g}\mathbb{P}\left(S_{z}\,|\,S_{z'},\bar{a}_{z'},\bar{a}_{z}\right).\label{eq:51}
\end{equation}
Define $\boldsymbol{a}^{z_{r}}$ as the strategy profile associated
with $z_{r}$ and define its objective by $J_{z_{r}}=\sum_{n=1}^{N}u_{n}\left(\boldsymbol{a}^{z_{r}}\right)$.
Throughout the proof we use the following definition.
\begin{defn}
Define $Z_{d}$ as the set of states for which exactly $d$ players
are discontent (so $N-d$ players are content). 
\end{defn}
We first provide an upper bound for \eqref{eq:49} by upper bounding
each factor in \eqref{eq:51} separately. We then construct a specific
tree $g\in G\left(z_{r}\right)$ that achieves the upper bound. 

\subsubsection{Upper bounding the probability of personal state transition - $\prod_{\left(z'\rightarrow z\right)\in g}\mathbb{P}\left(S_{z}\,|\,S_{z'},\bar{a}_{z'},\bar{a}_{z}\right)$ }

If $z_{r}\in Z_{0}$ then in the path from $z_{N}\in Z_{N}$ to $z_{r}$,
the personal state of each player needs to change from discontent
to content with $\boldsymbol{a}^{z_{r}}$. This occurs with a probability
of no more than $\left(\prod_{n=1}^{N}\frac{u_{n}\left(\boldsymbol{a}^{z_{r}}\right)}{u_{n,\max}}\right)\varepsilon^{\sum_{n}u_{n,\max}-J_{z_{r}}}$
since each transition to content with $u_{n}$ is only possible with
probability $\frac{u_{n}}{u_{n,\max}}\varepsilon^{u_{n,\max}-u_{n}}$
(see \eqref{eq:9}). From the second constraint on $G\left(z_{r}\right)$
(see Definition 5), this path exists in any tree in $G\left(z_{r}\right)$
such that $z_{r}\in Z_{0}$. We conclude that for any $z_{r}\in Z_{0}$
\begin{equation}
\prod_{\left(z'\rightarrow z\right)\in g}\mathbb{P}\left(S_{z}\,|\,S_{z'},\bar{a}_{z'},\bar{a}_{z}\right)\leq\varepsilon^{\sum_{n}u_{n,\max}-J_{z}}\prod_{n=1}^{N}\frac{u_{n}\left(\boldsymbol{a}^{z_{r}}\right)}{u_{n,\max}}.\label{eq:52}
\end{equation}
If $z_{r}\in Z_{d}$ for $d\geq1$ then trivially
\begin{equation}
\prod_{\left(z'\rightarrow z\right)\in g}\mathbb{P}\left(S_{z}\,|\,S_{z'},\bar{a}_{z'},\bar{a}_{z}\right)\leq1.\label{eq:53}
\end{equation}

\subsubsection{Upper bounding the probability of choosing a different arm - $\prod_{\left(z'\rightarrow z\right)\in g}\mathbb{P}\left(\bar{a}_{z}\,|\,S_{z'},\bar{a}_{z'}\right)$ }

Assume that $\varepsilon<\left(1-\frac{1}{M}\right)^{\frac{1}{c}}$
so $\frac{\varepsilon^{c}}{M-1}<1-\varepsilon^{c}$ and $1-\varepsilon^{c}>\frac{1}{M}$.
We use the first constraint on $G\left(z_{r}\right)$ (see Definition
5) and count the most likely outgoing edges from each non-root state. 
\begin{enumerate}
\item Any outgoing transition from $z'\in Z_{0}$ is bounded by the maximal
probability transition (in \eqref{eq:6}) 
\begin{equation}
\mathbb{P}\left(\bar{a}_{z}\,|\,S_{z'},\bar{a}_{z'},d=0\right)\leq\left(1-\varepsilon^{c}\right)^{N-1}\frac{\varepsilon^{c}}{M-1}\label{eq:54}
\end{equation}
even if the total state transition is infeasible. 
\item Any outgoing transition from $z'\in Z_{d}$ for $d\geq1$ is bounded
by the maximal probability transition (in \eqref{eq:6} and \eqref{eq:7})
\begin{equation}
\mathbb{P}\left(\bar{a}_{z}\,|\,S_{z'},\bar{a}_{z'},d\geq1\right)\leq\left(1-\varepsilon^{c}\right)^{N-d}\frac{1}{M^{d}}\label{eq:55}
\end{equation}
even if the total state transition is infeasible. 
\end{enumerate}
Since all $z\in Z\setminus\left\{ z_{r}\right\} $ must have a single
outgoing edge in $g$, going over all paths in $\left(z'\rightarrow z\right)\in g$
must at least include all these outgoing edges. By sorting these outgoing
edges according to the number of discontent players in their source
state, we conclude that for $z_{r}\in Z_{0}$ 
\begin{equation}
\prod_{\left(z'\rightarrow z\right)\in g}\mathbb{P}\left(\bar{a}_{z}\,|\,S_{z'},\bar{a}_{z'}\right)\leq\left(\frac{\left(1-\varepsilon^{c}\right)^{N-1}\varepsilon^{c}}{M-1}\right)^{\left|Z_{0}\right|-1}\prod_{d'=1}^{N}\left(\left(1-\varepsilon^{c}\right)^{N-d'}\frac{1}{M^{d'}}\right)^{\left|Z_{d'}\right|}\label{eq:56}
\end{equation}
and for $z_{r}\in Z_{d}$ for $d\geq1$ we have
\begin{equation}
\prod_{\left(z'\rightarrow z\right)\in g}\mathbb{P}\left(\bar{a}_{z}\,|\,S_{z'},\bar{a}_{z'}\right)\leq\frac{M^{d}}{\left(1-\varepsilon^{c}\right)^{N-d}}\left(\frac{\left(1-\varepsilon^{c}\right)^{N-1}\varepsilon^{c}}{M-1}\right)^{\left|Z_{0}\right|}\prod_{d'=1}^{N}\left(\left(1-\varepsilon^{c}\right)^{N-d'}\frac{1}{M^{d'}}\right)^{\left|Z_{d'}\right|}\label{eq:57}
\end{equation}
where the factor $\frac{M^{d}}{\left(1-\varepsilon^{c}\right)^{N-d}}$
compensates for counting one $Z_{d}$ state too many (the root). 

\subsubsection{Upper bounding $\prod_{\left(z'\rightarrow z\right)\in g}P_{z'z}$}

Define
\begin{equation}
A\left(\varepsilon,N,M\right)=\left(\frac{\left(1-\varepsilon^{c}\right)^{N-1}\varepsilon^{c}}{M-1}\right)^{\left|Z_{0}\right|-1}\prod_{d'=1}^{N}\left(\left(1-\varepsilon^{c}\right)^{N-d'}\frac{1}{M^{d'}}\right)^{\left|Z_{d'}\right|}.\label{eq:58}
\end{equation}
Gathering the terms from \eqref{eq:52},\eqref{eq:53},\eqref{eq:56},\eqref{eq:57}
together, we conclude that
\begin{equation}
\underset{g\in G\left(z_{r}\right)}{\max}\prod_{\left(z'\rightarrow z\right)\in g}P_{z'z}\leq\Biggl\{\begin{array}{cc}
\left(\prod_{n=1}^{N}\frac{u_{n}\left(\boldsymbol{a}^{z_{r}}\right)}{u_{n,\max}}\right)\varepsilon^{\sum_{n}u_{n,\max}-J_{z_{r}}}A\left(\varepsilon,N,M\right) & z_{r}\in Z_{0}\\
\left(1-\varepsilon^{c}\right)^{d-1}\frac{M^{d}}{M-1}\varepsilon^{c}A\left(\varepsilon,N,M\right) & z_{r}\in Z_{d},\,d\geq1
\end{array}.\label{eq:59}
\end{equation}

\subsubsection{Constructing a tree that achieves the bound in \eqref{eq:59} \label{subsec:Constructing-a-tree}}

\begin{figure}[tbh]
~~~~~~\includegraphics[width=8cm,height=8cm,keepaspectratio]{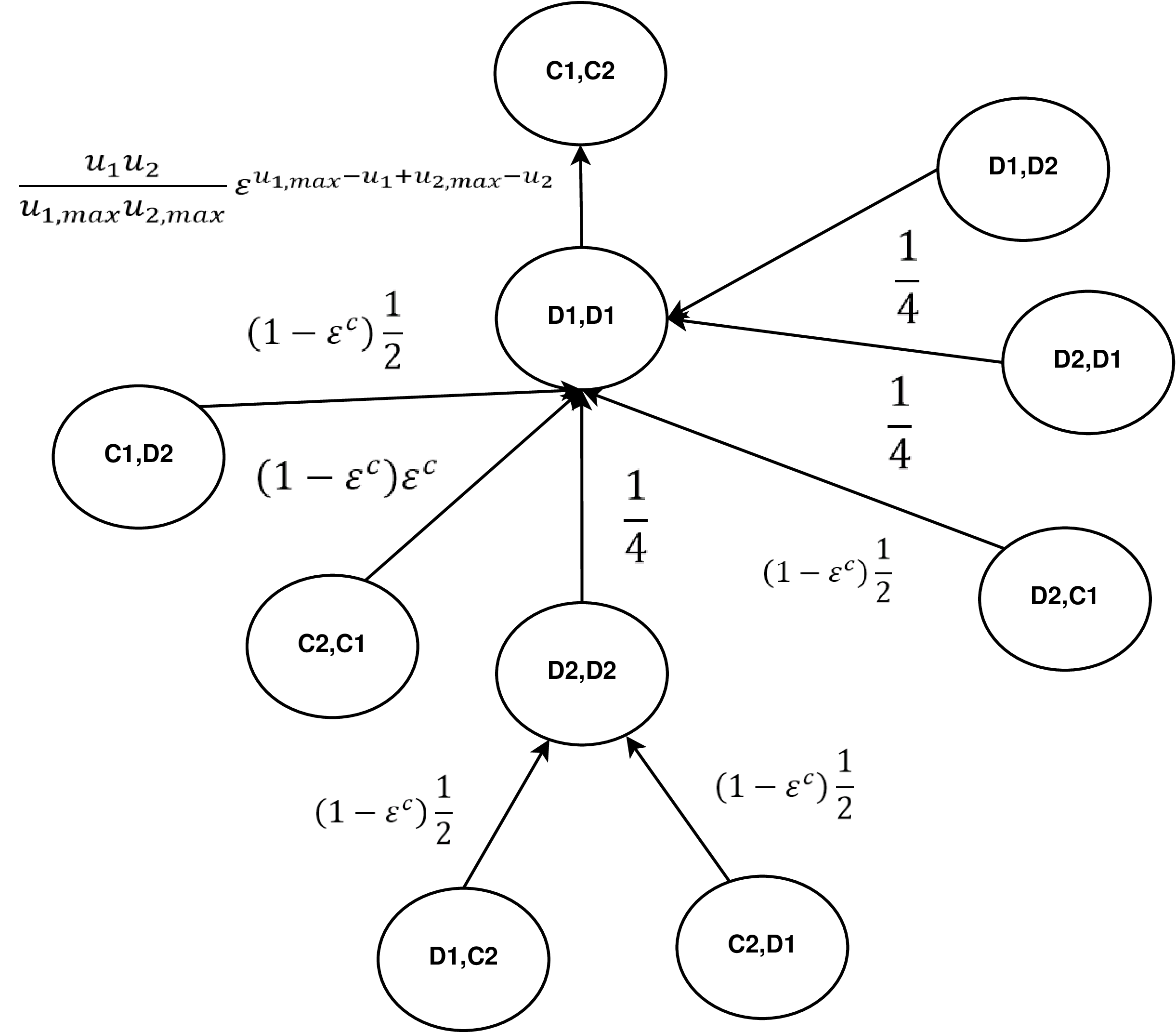}~~~~~\includegraphics[width=8cm,height=8cm,keepaspectratio]{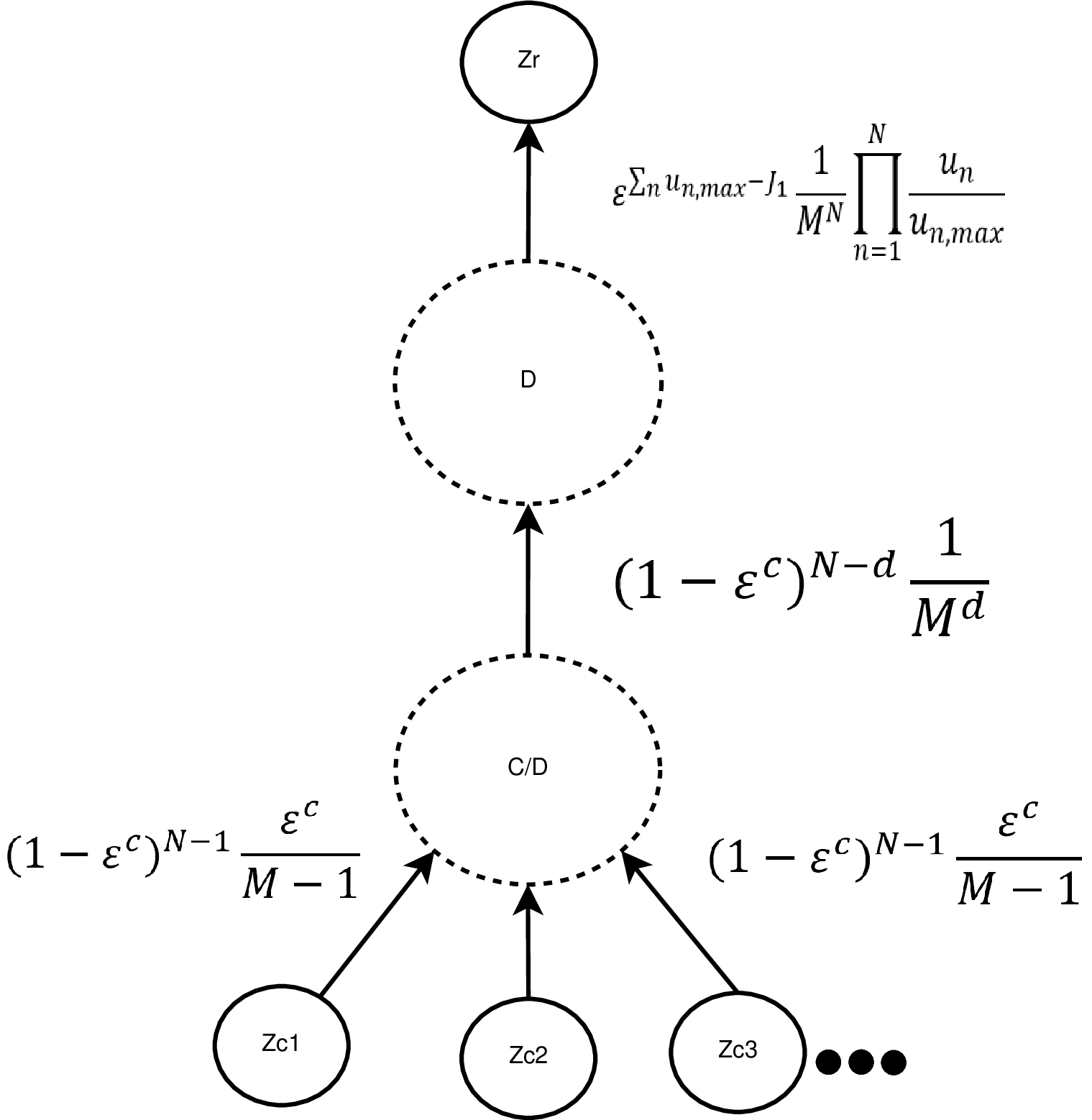}

\caption{The maximal tree of $z_{r}\in Z_{0}$ , for $N=M=2$ and for the general
case. C1,D2 is the state where player 1 is content with arm 1 and
player 2 is discontent with arm 2. \label{fig:Maximal Tree Two Players}}
\end{figure}

Now we construct a specific tree in $G\left(z_{r}\right)$ that achieves
the upper bound of \eqref{eq:59}, and hence must maximize \eqref{eq:49}.
We call this tree the maximal tree of $z_{r}$. This tree consists
solely of the maximal probability outgoing edges of each state. Fig.
\ref{fig:Maximal Tree Two Players} illustrates the maximal tree of
a content state $z_{0}\in Z_{0}$ for $N=M=2$. Note that the maximal
probability path between two states is not necessarily the shortest
one. For instance, there is a single edge in $Z$ from the lower content
state to the root, with a probability of $\varepsilon^{2c}$. However,
for $c>u_{\max,1}-u_{1}+u_{\max,2}-u_{2}$ and a small enough $\varepsilon$,
the probability of the most likely path is $\frac{\mu_{1}\mu_{2}}{u_{\max,1}u_{\max,2}}\left(1-\varepsilon^{c}\right)\varepsilon^{c}\varepsilon^{u_{\max,1}-u_{1}+u_{\max,2}-u_{2}}$. 

The maximal tree is constructed as follows:
\begin{enumerate}
\item Connect all $z_{0}\in Z_{0}$ to some state $z_{2}\in Z_{2}$ with
a probability of $\left(1-\varepsilon^{c}\right)^{N-1}\frac{\varepsilon^{c}}{M-1}$
. This is possible since the player that changed her arm (with a probability
of $\frac{\varepsilon^{c}}{M-1}$) chooses the arm of one of the other
players, making both of them discontent with probability 1 since they
both receive $u_{n}=0$. 
\item Connect all $z_{d}\in Z_{d}$ with $1\leq d<\frac{N}{2}$ to some
state $z_{d'}\in Z_{d'}$ with $d'>d$ with a probability of $\left(1-\varepsilon^{c}\right)^{N-d}\frac{1}{M^{d}}$.
This is possible when all the discontent players choose, with a probability
of $\frac{1}{M^{d}}$, one or more of the arms of the content players
(all kept their arms with a probability of $\left(1-\varepsilon^{c}\right)^{N-d}$),
thus making all the colliding players discontent with probability
1. 
\item Repeat Step 2 for $z_{d'}\in Z_{d'}$ until $d'\geq\frac{N}{2}$.
\item Connect all $z_{d}\in Z_{d}$ with $\frac{N}{2}\leq d<N$ to some
state $z_{N}\in Z_{N}$ with a probability of $\left(1-\varepsilon^{c}\right)^{N-d}\frac{1}{M^{d}}$.
This is possible since the discontent players together can choose
all the arms of the content players. Hence, after this transition,
all players receive $u_{n}=0$ and become discontent with probability
one.
\item Choose $\tilde{z}_{N}\in Z_{N}$ such that all players are in a collision.
Connect all other $z_{N}\in Z_{N}$ to $\tilde{z}_{N}$ with a probability
of $\frac{1}{M^{N}}$, recalling that the collision makes all players
discontent with probability 1. 
\item If $z_{r}\in Z_{N}$ then pick $\tilde{z}_{N}$ as the root. Otherwise
if $z_{r}\in Z_{0}$ then disconnect the outgoing edge of $z_{r}$
from step 1 and connect $\tilde{z}_{N}\in Z_{N}$ to $z_{r}$ with
a probability of $\frac{1}{M^{N}}\left(\prod_{n=1}^{N}\frac{u_{n}\left(\boldsymbol{a}^{z_{r}}\right)}{u_{n,\max}}\right)\varepsilon^{\sum_{n}u_{n,\max}-J_{z_{r}}}$.
\end{enumerate}
If the root is $z_{r}\in Z_{N}$ then we have constructed $g_{N}\in G\left(z_{r}\right)$
with
\begin{equation}
\prod_{\left(z'\rightarrow z\right)\in g_{N}}P_{z'z}=\left(1-\varepsilon^{c}\right)^{N-1}\frac{M^{N}}{M-1}\varepsilon^{c}A\left(\varepsilon,N,M\right).\label{eq:60}
\end{equation}
If the root is $z_{r}\in Z_{0}$, then we have constructed $\widetilde{g}_{N}\in G\left(z_{r}\right)$
with 
\begin{equation}
\prod_{\left(z'\rightarrow z\right)\in\widetilde{g}_{N}}P_{z'z}=\left(\prod_{n=1}^{N}\frac{u_{n}\left(a^{z_{r}}\right)}{u_{n,\max}}\right)\varepsilon^{\sum_{n}u_{n,\max}-J_{z_{r}}}A\left(\varepsilon,N,M\right).\label{eq:61}
\end{equation}
The maximal tree of $z_{r}$ is depicted in Fig. \ref{fig:Maximal Tree Two Players}.

\subsection{Upper bounding the probability of all non $z^{*}$ trees\label{subsec:rest of the trees}}

We are left with upper bounding $\sum_{z_{r}\in Z,z_{r}\neq z^{*}}\sum_{g\in G\left(z_{r}\right)}\prod_{\left(z'\rightarrow z\right)\in g}P_{z'z}$.
First we identify a probability factor that appears in $\prod_{\left(z'\rightarrow z\right)\in g}P_{z'z}$
for every tree $g$. In any possible outgoing edge from $z_{d}\in Z_{d}$
there is a probability of $\frac{1}{M^{d}}$ for the action choices
of the $d$ discontent players. In a similar manner, there is a factor
of $\frac{\varepsilon^{c}}{M-1}$ in the probability of any possible
outgoing edge from any $z_{0}\in Z_{0}$. Hence, any tree has
\begin{equation}
B\left(\varepsilon,N,M\right)=\left(\frac{\varepsilon^{c}}{M-1}\right)^{\left|Z_{0}\right|-1}\prod_{d'=1}^{N}\left(\frac{1}{M^{d'}}\right)^{\left|Z_{d'}\right|}\label{eq:63}
\end{equation}
as a factor of $\prod_{\left(z'\rightarrow z\right)\in g}P_{z'z}$.
Moreover:
\begin{enumerate}
\item All trees $g\in G\left(z_{r}\right)$ with $z_{r}\in Z_{0}$ have
a factor of $\prod_{n=1}^{N}\frac{u_{n}\left(\boldsymbol{a}^{z_{r}}\right)}{u_{n,\max}}\varepsilon^{\sum_{n}u_{n,\max}-J_{z_{r}}}$
in $\prod_{\left(z'\rightarrow z\right)\in g}P_{z'z}$ since there
must exist a path from any $z_{N}\in Z_{N}$ to the root, and all
players change from being discontent to being content with $\boldsymbol{a}_{z_{r}}$
at least once on this path. 
\item All trees $g\in G\left(z_{r}\right)$ with $z_{r}\notin Z_{0}$ must
have a factor of $\left(\frac{\varepsilon^{c}}{M-1}\right)^{\left|Z_{0}\right|}$
(instead of $\left(\frac{\varepsilon^{c}}{M-1}\right)^{\left|Z_{0}\right|-1}$)
in $\prod_{\left(z'\rightarrow z\right)\in g}P_{z'z}$ since there
are $\left|Z_{0}\right|$ nodes from $Z_{0}$ to be connected.
\end{enumerate}
Now we divide all the trees in $G\left(z\right)$ into different categories.
Define $a_{l,q}^{z}$ as the number of trees in $G\left(z\right)$
that have an $l$ extra $\varepsilon^{u_{n,\max}-u_{n}}<1$ factors
 and $q$ extra $\varepsilon^{c}$ factors, where extra means more
than the factors that must exist in any tree as argued above. Note
that since there are $N$ players and $\left|Z\right|-1$ edges in
every tree, $l$ and $q$ cannot exceed $N\left|Z\right|$. Let $\mathcal{A}{}_{\max}^{n}=\arg\underset{\boldsymbol{a}}{\max}\,u_{n}\left(\boldsymbol{a}\right)$
and define
\begin{equation}
\alpha=\underset{n}{\min}\underset{\boldsymbol{a}\notin\mathcal{A}{}_{\max}^{n}}{\min}\left(u_{n,\max}-u_{n}\left(\boldsymbol{a}\right)\right).\label{eq:64}
\end{equation}
Also define $J_{0}\triangleq\sum_{n}u_{n,\max}$ and recall Definition
3 of the second best objective $J_{2}$. Using the above definitions,
we can write 
\begin{multline}
\sum_{z_{r}\in Z,z\neq z^{*}}\sum_{g\in G\left(z_{r}\right)}\prod_{\left(z'\rightarrow z\right)\in g}P_{z'z}=B\left(\varepsilon,N,M\right)\frac{\varepsilon^{c}}{M-1}\sum_{d=1}^{N}\sum_{z_{r}\in Z_{d}}\sum_{g\in G\left(z_{r}\right)}\frac{\prod_{\left(z'\rightarrow z\right)\in g}P_{z'z}}{\frac{\varepsilon^{c}}{M-1}B\left(\varepsilon,N,M\right)}+\\
B\left(\varepsilon,N,M\right)\sum_{z_{r}\in Z_{0},z\neq z^{*}}\varepsilon^{J_{0}-J_{z_{r}}}\prod_{n=1}^{N}\frac{u_{n}\left(\boldsymbol{a}^{z_{r}}\right)}{u_{n,\max}}\sum_{g\in G\left(z_{r}\right)}\frac{\prod_{\left(z'\rightarrow z\right)\in g}P_{z'z}}{\varepsilon^{J_{0}-J_{z_{r}}}\prod_{n=1}^{N}\frac{u_{n}\left(\boldsymbol{a}^{z_{r}}\right)}{u_{n,\max}}B\left(\varepsilon,N,M\right)}\underset{\left(a\right)}{\leq}\\
B\left(\varepsilon,N,M\right)\left(\frac{\varepsilon^{c}}{M-1}\sum_{d=1}^{N}M^{d}\sum_{z_{r}\in Z_{d}}\sum_{l=0}^{N\left|Z\right|}\sum_{q=0}^{N\left|Z\right|}a_{l,q}^{z_{r}}\varepsilon^{l\alpha}\varepsilon^{qc}+\sum_{z_{r}\in Z_{0},z\neq z^{*}}\varepsilon^{J_{0}-J_{z_{r}}}\prod_{n=1}^{N}\frac{u_{n}\left(\boldsymbol{a}^{z_{r}}\right)}{u_{n,\max}}\sum_{l=0}^{N\left|Z\right|}\sum_{q=0}^{N\left|Z\right|}a_{l,q}^{z_{r}}\varepsilon^{l\alpha}\varepsilon^{qc}\right)\\
\underset{\left(b\right)}{\leq}\frac{B\left(\varepsilon,N,M\right)}{\left(1-\varepsilon^{\alpha}\right)\left(1-\varepsilon^{c}\right)}\left(\frac{\varepsilon^{c}}{M-1}\sum_{d=1}^{N}M^{d}\sum_{z_{r}\in Z_{d}}\underset{l,q}{\max}a_{l,q}^{z_{r}}+\sum_{z_{r}\in Z_{0},z\neq z^{*}}\underset{l,q}{\max}a_{l,q}^{z_{r}}\prod_{n=1}^{N}\frac{u_{n}\left(\boldsymbol{a}^{z_{r}}\right)}{u_{n,\max}}\varepsilon^{J_{0}-J_{z_{r}}}\right)\leq\\
\frac{B\left(\varepsilon,N,M\right)}{\left(1-\varepsilon^{\alpha}\right)\left(1-\varepsilon^{c}\right)}\left(\left|Z\right|\underset{z_{r},l,q}{\max}a_{l,q}^{z_{r}}\right)\left(\frac{M\left(M^{N}-1\right)}{\left(M-1\right)^{2}}\varepsilon^{c}+\varepsilon^{J_{0}-J_{2}}\right)\label{eq:65}
\end{multline}
where (a) follows since $B\left(\varepsilon,N,M\right)$ and $\frac{\varepsilon^{c}}{M-1}$
or $\varepsilon^{J_{0}-J_{z_{r}}}\prod_{n=1}^{N}\frac{u_{n}\left(\boldsymbol{a}^{z_{r}}\right)}{u_{n,\max}}$
are factors of $\prod_{\left(z'\rightarrow z\right)\in g}P_{z'z}$,
and so are $\varepsilon^{ld}\varepsilon^{qc}$ if $g$ is counted
by $a_{l,q}^{z}$. The factor $M^{d}$ compensates for counting, in
\eqref{eq:63}, the (non-existing) outgoing edge of $z_{r}\in Z_{d}$.
Inequality (b) follows from the two geometric series that appear by
bounding as follows
\begin{equation}
\sum_{l=0}^{N\left|Z\right|}\sum_{q=0}^{N\left|Z\right|}a_{l,q}^{z_{r}}\varepsilon^{l\alpha}\varepsilon^{qc}\leq\underset{l,q}{\max}a_{l,q}^{z_{r}}\sum_{l=0}^{\infty}\sum_{q=0}^{\infty}\varepsilon^{l\alpha}\varepsilon^{qc}=\frac{1}{1-\varepsilon^{\alpha}}\frac{1}{1-\varepsilon^{c}}\underset{l,q}{\max}a_{l,q}^{z_{r}}.\label{eq:66}
\end{equation}

\subsection{Lower bounding $\pi_{z^{*}}$}

Using \eqref{eq:65} we obtain
\begin{multline}
\frac{\sum_{z_{r}\in Z,z\neq z^{*}}\sum_{g\in G\left(z_{r}\right)}\prod_{\left(z'\rightarrow z\right)\in g}P_{z'z}}{\sum_{g\in G\left(z^{*}\right)}\prod_{\left(z'\rightarrow z\right)\in g}P_{z'z}}\underset{\left(a\right)}{\leq}\frac{\frac{B\left(\varepsilon,N,M\right)}{\left(1-\varepsilon^{\alpha}\right)\left(1-\varepsilon^{c}\right)}\left(\left|Z\right|\underset{z_{r},l,q}{\max}\,a_{l,q}^{z_{r}}\right)\left(\frac{M\left(M^{N}-1\right)}{\left(M-1\right)^{2}}\varepsilon^{c}+\varepsilon^{J_{0}-J_{2}}\right)}{G^{*}\left(\prod_{n=1}^{N}\frac{u_{n}\left(\boldsymbol{a}^{*}\right)}{u_{n,\max}}\right)\varepsilon^{J_{0}-J_{1}}A\left(\varepsilon,N,M\right)}=\\
\frac{\prod_{n=1}^{N}\frac{u_{n,\max}}{u_{n}\left(\boldsymbol{a}^{*}\right)}}{\left(1-\varepsilon^{\alpha}\right)\left(1-\varepsilon^{c}\right)}\frac{\left(\frac{\left|Z\right|}{G^{*}}\underset{z_{r},l,q}{\max}\,a_{l,q}^{z_{r}}\right)\left(\frac{M\left(M^{N}-1\right)}{\left(M-1\right)^{2}}\varepsilon^{c-J_{0}+J_{1}}+\varepsilon^{J_{1}-J_{2}}\right)}{\left(1-\varepsilon^{c}\right)^{\left(N-1\right)\left(\left|Z_{0}\right|-1\right)}\prod_{d'=1}^{N}\left(1-\varepsilon^{c}\right)^{\left(N-d'\right)\left|Z_{d'}\right|}}\underset{\left(b\right)}{\leq}\\
\frac{\prod_{n=1}^{N}\frac{u_{n,\max}}{u_{n}\left(\boldsymbol{a}^{*}\right)}}{\left(1-\varepsilon^{\alpha}\right)\left(1-\varepsilon^{c}\right)^{\left(N-1\right)2^{N}M^{N}}}\left(2^{N}M^{N}\frac{\underset{z_{r},l,q}{\max}a_{l,q}^{z_{r}}}{G^{*}}\right)\left(3M^{N-1}\varepsilon^{c-J_{0}+J_{1}}+\varepsilon^{J_{1}-J_{2}}\right)\label{eq:67}
\end{multline}
where in (a) we used \eqref{eq:65} for the numerator, where $G^{*}$
denotes the number of maximal trees of $z^{*}$ (from Subsection \ref{subsec:Constructing-a-tree})
for the denominator. Inequality (b) follows since $\sum_{d'=1}^{N}\left(N-d'\right)\left|Z_{d'}\right|\leq\left(N-1\right)\left(\left|Z\right|-\left|Z_{0}\right|\right)$
and $\left|Z\right|=2^{N}M^{N}$. We also used $\frac{M\left(M^{N}-1\right)}{\left(M-1\right)^{2}}\leq3M^{N-1}$,
which holds for all $M>1$.

We conclude that if $c>\sum_{n}u_{n,\max}-J_{1}$ then \eqref{eq:67}
vanishes to zero as $\varepsilon\rightarrow0$. Hence, $z^{*}$ is
the most likely state in the stationary distribution of $Z$ for small
enough $\varepsilon$. Specifically, having 
\begin{equation}
\varepsilon<\min\left\{ \left(\prod_{n=1}^{N}\frac{u_{n}\left(\boldsymbol{a}^{*}\right)}{u_{n,\max}}\left(2M\right)^{-N}\frac{3G^{*}}{8\underset{z_{r},l,q}{\max}a_{l,q}^{z_{r}}}\right)^{\frac{1}{J_{1}-J_{2}}},\left(\prod_{n=1}^{N}\frac{u_{n}\left(\boldsymbol{a}^{*}\right)}{u_{n,\max}}\left(2M\right)^{-N}M^{-N+1}\frac{G^{*}}{8\underset{z_{r},l,q}{\max}a_{l,q}^{z_{r}}}\right)^{\frac{1}{c-J_{0}+J_{1}}}\right\} \label{eq:68}
\end{equation}
together with 
\begin{equation}
\varepsilon<\min\left\{ \frac{1}{10^{1/\alpha}},\,\left(1-\left(\frac{9}{10}\right)^{\frac{1}{\left(N-1\right)2^{N}M^{N}}}\right)^{1/c}\right\} \label{eq:69}
\end{equation}
 is enough to ensure that $\pi_{z^{*}}>\frac{1}{2}$. Note that \eqref{eq:68}
is typically much stricter than \eqref{eq:69}. Also, $\prod_{n=1}^{N}\frac{u_{n}\left(\boldsymbol{a}^{*}\right)}{u_{n,\max}}$
is typically not much below one. Evaluating the factor $\frac{G^{*}}{\underset{z_{r},l,q}{\max}a_{l,q}^{z_{r}}}$
involves counting the number of trees in each category and can be
done by a computer program. \bibliographystyle{IEEEtran}
\bibliography{GoT}

\end{document}